\documentclass{elsart}

    \usepackage{latexsym,bm,amsmath,amssymb,amsfonts}
    \usepackage{epsfig,graphics,graphicx}
    \usepackage{makeidx}
    \usepackage{citesort}

    \newcommand{\dif}{{\rm d}}
    
    \newcommand{\del}{\partial}
    
    \newcommand{\dA}{\delta\!A}
    \newcommand{\wt}[1]{\widetilde{#1}}

\long\def\comment#1{ }

\newcommand{\nn}{\nonumber\\ }
\newcommand{\beq}{\begin{eqnarray}}
\newcommand{\eeq}{\end{eqnarray}}
\newcommand{\be}{\vspace{-.4cm}\begin{eqnarray}}
\newcommand{\ee}{\vspace{-.5cm}\end{eqnarray}}

\def\labe{\label}
\newcommand{\lan}{\langle}
\newcommand{\ran}{\rangle}
\def\grad{\nabla}
\def\del{\partial}

\newcommand{\cal}{\mathcal} 
\newcommand{\rme}{{\rm e}}
\newcommand{\tr}{{\rm tr}}
\newcommand{\Tr}{{\rm Tr}}

\def\simge{\mathrel{%
   \rlap{\raise 0.511ex \hbox{$>$}}{\lower 0.511ex \hbox{$\sim$}}}}
\def\simle{\mathrel{
   \rlap{\raise 0.511ex \hbox{$<$}}{\lower 0.511ex \hbox{$\sim$}}}}

\newcommand{\x}{\bm x}
\newcommand{\y}{\bm y}

\newcommand{\z}{\bm z}

\begin{document}

\begin{flushright}
~\vspace{-1.25cm}\\
{\small\sf SACLAY--T05/057\\
 BNL--NT--05/11}
\end{flushright}
\vspace{2.cm}

\begin{frontmatter}

\parbox[]{16.0cm}{ \begin{center}
\title{Effective Hamiltonian for QCD evolution at high energy}

\author{Y.~Hatta$^{\rm a}$},
\author{E.~Iancu$^{\rm b,}$\thanksref{th2}},
\author{L.~McLerran$^{\rm a,c}$},
\author{A.~Sta\'sto$^{\rm c,d}$},
\author{D.N.~Triantafyllopoulos$^{\rm b}$}

\address{$^{\rm a}$ RIKEN BNL Research Center, Brookhaven National Laboratory,
Upton, NY 11973, USA}

\address{$^{\rm b}$Service de Physique Theorique, Saclay,
F-91191 Gif-sur-Yvette, France}

\address{$^{\rm c}$ Physics Department, Brookhaven
National Laboratory, Upton, NY 11973, USA}

\address{$^{\rm d}$ Institute of Nuclear Physics PAN,
Radzikowskiego 152, Krak\'ow, Poland}

\thanks[th2]{Membre du Centre National de la Recherche Scientifique
(CNRS), France.}

\date{\today}
\vspace{0.8cm}
\begin{abstract}
We construct the effective Hamiltonian which governs the
renormalization group flow of the gluon distribution with increasing
energy and in the leading logarithmic approximation. This Hamiltonian
defines a two--dimensional field theory which involves two types of
Wilson lines: longitudinal Wilson lines which describe gluon
recombination (or merging) and temporal Wilson lines which account for
gluon bremsstrahlung (or splitting). The Hamiltonian is self--dual,
i.e., it is invariant under the exchange of the two types of Wilson
lines. In the high density regime where one can neglect gluon number
fluctuations, the general Hamiltonian reduces to that for the JIMWLK
evolution. In the dilute regime where gluon recombination becomes
unimportant, it reduces to the dual partner of the JIMWLK Hamiltonian,
which describes bremsstrahlung.

\end{abstract}
\end{center}}

\end{frontmatter}
\newpage

\section{Introduction}
\setcounter{equation}{0}

The construction of an effective field theory for scattering in QCD at
high energy is a difficult, longstanding, problem, which has been at
the heart of theoretical developments for more than twenty years. In
the `leading logarithmic approximation' (LLA) which is expected to
control the high--energy limit, the effective action should include the
quantum effects enhanced by the large energy logarithm $\tau\equiv\ln
s/Q^2$ (the `rapidity'; as usual, $s$ denotes the invariant total
energy squared, and $Q$ is the typical momentum transfer, or a particle
mass), that is, it must resum radiative corrections of order $(\alpha_s
\ln s/Q^2)^n$ to all orders. If this is the only requirement, then the
corresponding resummation is performed by the
Balitsky--Fadin--Kuraev--Lipatov (BFKL) equation \cite{BFKL}, which is
however well known to violate unitarity in the high energy limit: the
corresponding solution for the scattering amplitudes or for the gluon
distribution grows like a power of the energy, in violation of the
Froissart bound. According to the modern understanding, the BFKL
equation should govern only the pre--asymptotic evolution, at
intermediate values of the energy. The complete effective action valid
in the high energy limit should unitarize the BFKL `pomeron' via the
inclusion of multiple scattering. Alternatively, and equivalently, if
the effective theory is written for the wavefunctions of the hadronic
systems which participate in the collision (the `target' and the
`projectile'), the effective action must include the non--linear
effects, like gluon recombination, responsible for {\em gluon
saturation} \cite{GLR,MQ85,BM87,MV}.

The first attempt towards constructing such an action has been
given in Ref. \cite{VV93}, where the main arguments rely on
kinematical and symmetry considerations. A lasting conclusion of
this analysis is that the effective action should describe a
two--dimensional field theory living in the plane transverse to
the collision axis, and that the basic degrees of freedom should
be {\em Wilson line} --- path--ordered eikonal phases describing
multiple scattering at high energy. The effective theory proposed
in Ref. \cite{VV93} involves two types of such Wilson lines, each
of them describing the scattering of one of the participants in
the collision off the color field created by the other
participant. Subsequently, Lipatov and collaborators
\cite{LKSeff,LipatovS} proposed an effective theory formulated in
terms of reggeized gluons which contains the results in Ref.
\cite{VV93} as a special limit. More recently, Balitsky \cite{B1}
has relied on a new factorization scheme (in rapidity) to
construct an effective action in terms of Wilson lines, which
extends the original construction in Ref. \cite{VV93}. Some of his
results will also emerge from our subsequent calculations. The
direct comparison between the effective theory by Lipatov et al
\cite{LKSeff,LipatovS} and that by Balitsky (or, more generally,
any effective theory built in terms of Wilson lines, like the one
to be developed below in this paper) is hindered by the fact that
the general relation between the reggeized gluons and the usual
gluon fields (and hence the Wilson lines) is not known.

However, all the approaches mentioned above have merely considered
the effective action corresponding to a {\em given layer in
rapidity} : The action is obtained by integrating out gluon
fluctuations with rapidity $\tau$ within some intermediate range
$\tau_0 < \tau < \tau_1$, to leading order in
$\alpha_s(\tau_1-\tau_0)$ and under the assumption that
$\alpha_s(\tau_1-\tau_0)\ll 1$, in order for perturbation theory
to apply. (The rapidity of a parton fluctuation is defined, as
usual, as $\tau=\ln 1/x$, where $x$ is the longitudinal momentum
fraction carried by that parton.) From the point of view of the
leading logarithmic approximation, this represents a {\em single
evolution step} (in the presence of multiple scattering, though),
but in order to study the high energy behavior one really needs to
{\em iterate} this step. To that aim, one should be able to
promote the effective action into a (field theoretical) {\em
renormalization group Hamiltonian} which should describe the
evolution of the effective theory with increasing $\tau$. In turn,
this requires the proper identification of the relevant degrees of
freedom and of the way that these are changed by the evolution in
each rapidity step.

Alternatively, the  renormalization group (RG) flow can be
formulated as an hierarchy of evolution equations for the relevant
gluon correlation functions, which should reduce to the BFKL
equation in the intermediate--energy regime where unitarity
corrections are unimportant. Such an hierarchy has been derived
for the first time by Balitsky \cite{B}. This is an hierarchy for
the evolution of Wilson--line operators which physically describe
the scattering between a projectile made of elementary color
charges (each of them identified by its own Wilson line) and the
strong `background' field of a target which is not evolving (like
a large nucleus). The evolution consists in gluon splitting within
the projectile wavefunction, followed by the multiple scattering
between the products of this splitting and the background field.
With increasing $\tau$, the gluon density in the projectile
wavefunction is rapidly increasing, yet this formalism does not
allow for non--linear effects like gluon recombination, which on
physical grounds are expected to tame this growth.

The interpretation of Balitsky equations as an RG flow has been
clarified by Weigert \cite{W}, who noticed that the hierarchy in
Ref. \cite{B} can be compactly reformulated as a single,
functional, evolution equation for a classical probability
distribution for the Wilson lines. The functional differential
operator in this equation plays the role of a Hamiltonian; it
involves only one type of Wilson line (since the scattering is
asymmetric: the target is dense and produces a strong color field,
while the projectile is dilute), together with functional (Lie)
derivatives with respect to these Wilson lines. {\em A
posteriori}, one can recognize Weigert's Hamiltonian as a {\em
`quantized'} version of the effective action proposed by Balitsky
for asymmetric scattering \cite{B}, where by `quantization' we
mean that the weak fields associated in the effective action with
the projectile are replaced by functional Lie derivatives with
respect to the Wilson lines built with the field of the target.

Independently, a different RG approach aiming at the description of the
gluon distribution of a highly evolved hadronic system has been
developed in Refs. \cite{MV,K96,JKMW97,JKLW97,RGE} and has eventually
crystalized into an effective theory for gluon correlations at small
$x$ known as the {\em color glass condensate} (CGC) \cite{RGE} (see
also the review papers \cite{CGCreviews}). Following the general idea
of the separation of scales in rapidity, this approach focuses on the
evolution of a single hadron wavefunction (say, the target), with the
purpose of integrating out the `fast' quantum fluctuations --- those
with large longitudinal momenta
--- in layers of rapidity and thus obtain an effective theory for the
correlations of the `soft' (i.e., small--$x$) gluons, as probed in a
high energy collision. The Wilson lines appear naturally also in this
approach: They describe the multiple scattering between the quantum
gluons to be integrated out in the next evolution step and the strong
color fields generated by `color sources' representing the fast gluons
which have been integrated out in the previous steps.

The central result in the CGC approach is the
Jalilian-Marian--Iancu--McLerran--Weigert--Leonidov--Kovner (JIMWLK)
equation \cite{JKLW97,RGE}, a functional RG equation which turns out to
be equivalent \cite{RGE,PATH} with the equation deduced by Weigert
\cite{W} from the Balitsky hierarchy \cite{B}. This equivalence
reflects the fact that both approaches include the same physical
non--linear phenomena, which are treated either as recombination
effects in the target wavefunction (in the CGC approach) or as gluon
splitting in the projectile wavefunction followed by multiple
scattering off the target (in the approach by Balitsky). However, both
approaches miss the other side of the problem, namely the saturation
effects in the evolution of the projectile (for the Balitsky equations)
and, respectively, the correlations associated with gluon splitting in
the target wavefunction (in the case of the JIMWLK equation). Because
of that, both approaches miss the `pomeron loops', as first recognized
in Ref. \cite{IT04}. These are the loops which open up with vertices
for gluon splitting (so like $2\to n$ with $n\ge 3$) in the dilute
regime and close through vertices for gluon merging (e.g., $n\to 2$) in
the high density regime. (The denomination `pomeron loops' is strictly
correct only in the limit where the number of colors $N_c$ is large and
the $t$--channel gluons can be pairwise combined into BFKL pomerons,
but here we use this expression as a convenient short--cut also for the
general case at arbitrary $N_c$.)

The gluon splitting and merging vertices in the high--energy
evolution are generalizations of the $2\to 2$ BFKL vertex in the
Lipatov Hamiltonian \cite{LipatovH}. Such vertices have been
explicitly computed in perturbative QCD
\cite{BW93,BE99,BV99,ES04,BBV05} (see \cite{ewerz} for a review)
and appear implicitly as building blocks in the Balitsky and
JIMWLK equations, from which they can be in principle extracted by
expanding the Wilson lines in powers of the gauge fields. However,
until recently the only formalism allowing for the simultaneous
inclusion of both splittings and mergings in a high--energy
scattering was Mueller's center--of--mass factorization of
onium--onium scattering \cite{AM94,AM95,IM031}. In this approach,
the gluon splittings are included in the evolution of the
wavefunctions of the two onia (described as evolving distributions
of quark--antiquark `color dipoles', as appropriate at large
$N_c$), while the mergings are represented by multiple scattering
(the simultaneous scattering of several pairs of dipoles from the
two onia). The numerical simulations of the onium--onium
scattering by Salam \cite{Salam95} have demonstrated the
importance of the correlations induced by gluon splitting, whose
physical role has been fully understood only recently
\cite{IM032,MS04,IMM04}: the gluon number fluctuations act as a
seed for higher--point correlations in the dilute regime, and thus
strongly influence the evolution towards high density.

The first evolution equations in QCD to take into account both mergings
and splittings have been proposed in Ref. \cite{IT04} and shortly after
improved in Refs. \cite{MSW05,IT05}. These equations generalize
Balitsky equations at large $N_c$ by including the effects of gluon
number fluctuations within the framework of the color dipole picture.
By construction, these equations describe dipole evolution (including
dipole number fluctuations) in the dilute regime and JIMWLK--like gluon
recombination in the high--density regime, and thus contain all the
necessary ingredients to provide a complete picture of the evolution at
large $N_c$ and for sufficiently high energies. The RG formulation of
these equations is also explicitly known \cite{MSW05,BIIT05}: this
involves an effective Hamiltonian which describes a two--dimensional
field theory for the dynamics of `Pomerons' \cite{BIIT05}. In this
context, the `Pomerons' represent color singlet exchanges in the
scattering between a dipole and a generic target. With increasing
$\tau$, the Pomerons undergo BFKL evolution, they can dissociate (one
Pomeron splitting into two) or recombine with each other (two Pomerons
merging into one). This effective theory has a remarkable symmetry
which reflects boost invariance: It is invariant under the {\em duality
transformations} which exchange color fields with the functional
derivative with respect to these fields (or, more precisely, with
respect to their sources). Under these transformations, the splitting
and mergings terms in the Hamiltonian get interchanged with each other,
while the BFKL piece is self--dual.

Whereas for large $N_c$, the high--energy evolution in QCD is by
now well understood \cite{IT04,IT05,MSW05,LL05,BIIT05,Levin05},
its generalization to finite $N_c$ remains an outstanding open
problem, which is actively under investigation
\cite{KL05,KL3,KL4}. Recently, Kovner and Lublinsky \cite{KL05}
have constructed evolution equations for gluon correlations in the
dilute regime valid for arbitrary $N_c\,$, and then noticed
\cite{KL3} that the effective Hamiltonian generating these
equations --- to which we shall refer as the {\em bremsstrahlung}
(BREM) {\em Hamiltonian} in what follows --- is in fact dual to
the JIMWLK Hamiltonian \cite{W,RGE}. Following this idea, they
made the interesting suggestion \cite{KL3} that the general (and
yet unknown) evolution Hamiltonian, which describes both mergings
and splittings, should be {\em self--dual} as a consequence of
boost invariance. But in the absence of an explicit derivation of
this Hamiltonian from the QCD Feynman rules, the hypotheses in
Ref. \cite{KL3} are difficult to test. Besides, by itself, the
duality argument does not allow one to construct the general
Hamiltonian for the QCD evolution, but only to relate its limiting
expressions in the dense and dilute regimes.

In fact, related work by the same authors \cite{KL05,KL4} shows that
the extension of the JIMWLK equation to the dilute regime is afflicted
with difficulties associated, in particular, with ambiguities due to
the time--ordering of the correlations induced by quantum evolution. At
this point, one should recall that previous formalisms like the CGC
approach and the Balitsky equations, and also the effective action
approaches in Refs. \cite{VV93,LKSeff,LipatovS,B1}, have been specially
tailored to deal with the high density, or strong field, regime, as
relevant in the vicinity of the unitarity limit, but they need not be
well suited for the description of subtle correlations induced by
fluctuations in the dilute regime. Thus, in view of the current
understanding, the very existence of a general evolution Hamiltonian
which encompasses all the desired phenomena is far from being clear,
and the actual expression of this Hamiltonian, if any, is completely
unknown.

It is our main objective in this paper to investigate under which
conditions the renormalization group approach developed in Refs.
\cite{JKLW97,RGE} can be extended to account for gluon splitting in
addition to gluon recombination, and thus promoted into a general
theory for QCD evolution in the LLA. As we shall see, such an extension
can be given indeed, but only within the limitations inherent to an RG
analysis, which implies a coarse--graining as usual. Here, the
`coarse--graining' refers, of course, to rapidity, and thus to the
longitudinal and temporal scales simultaneously. The effective theory
can capture only those correlations that can be `seen' by the quantum
fluctuations which are integrated out in one step of the evolution. As
we shall briefly explain here (and then demonstrate at length in the
main body of the paper), such correlations are fully encoded in {\em
two types of Wilson lines} --- path--ordered exponentials in the
longitudinal and, respectively, temporal direction
--- which with increasing $\tau$ evolve through gauge rotations
at their {\em end points}.

Indeed, one step of quantum evolution consists in integrating out
`semi--fast' gluon fluctuations in a rapidity layer $\tau_0 < \tau <
\tau_1$ and in the presence of two types of background fields: the
`fast' fields created by the gluons with larger longitudinal momenta
that have been integrated out in the previous steps, and the `slow'
fields radiated by the `semi--fast' gluon itself. The multiple
scattering off the `fast' background fields accounts for gluon merging,
whereas the simultaneous emission of several `slow' fields implements
gluon splitting (or bremsstrahlung). The resolution scales of the
semi--fast gluon are fixed by the light--cone kinematics together with
the separation of scales: This gluon has a poor resolution both on the
longitudinal scale characteristic for the fast fields and on the
temporal scale characteristic for the slow ones. Thus, it can only
measure an `integrated' effect of these fields, and since this measure
proceeds via multiple scattering, its result is encoded in Wilson
lines. The supports of the longitudinal and temporal integrations
implicit in the Wilson lines are evolving with $\tau$, but the
background  fields change only at the extremities of these supports, so
the Wilson lines change only at their end points (where new
correlations are induced by the quantum evolution).

The effective Hamiltonian that we shall obtain is a functional of
these two types of Wilson lines, and is moreover {\em
self--dual\,}: that is, it is invariant under the exchange between
longitudinal and temporal Wilson lines. This invariance reflects a
mirror symmetry between vertices for gluon splitting and,
respectively, gluon merging, which is in turn related to the boost
invariance of the evolution equations. In the limiting regimes
where one of the background fields is weak (so that the
corresponding Wilson lines can be expanded out in perturbation
theory), the general Hamiltonian reduces to the expected limiting
forms: \texttt{(i)} at high density, where the gluon number
fluctuations become negligible, it reduces to the JIMWLK
Hamiltonian \cite{W,RGE}; \texttt{(ii)} at low density, where
gluon recombination can be ignored, it reduces to the
bremsstrahlung Hamiltonian \cite{KL3}, which is dual to the JIMWLK
Hamiltonian, in agreement with Ref. \cite{KL3}.

For each of the two limiting situations alluded to above (JIMWLK
and BREM), the two--dimensional Hamiltonian structure is explicit
and transparent: the canonical variables and the associated
conjugate momenta are properly identified, and the corresponding
Poisson brackets are explicitly known. Note that these Poisson
brackets are non--trivial, in the sense that some of the variables
--- those associated with the weak fields and which play the role
of Lie derivatives --- are non--commutative: they do not commute
with themselves, but rather obey the SU($N_c$) current algebra.
The evolution equations are then obtained as the canonical
equations of motion generated by the respective Hamiltonian via
the Poisson brackets. For the JIMWLK Hamiltonian this procedure
yields the Balitsky equations \cite{B}, as usual, whereas for the
BREM Hamiltonian it leads to equations similar to those written
down by Kovner and Lublinsky \cite{KL05}.

But precisely because of the lack of commutativity alluded to
above, we have not been able to extend the construction of the
Poisson brackets to the general case, where both types of
background fields are strong. The would--be straightforward
construction, which consists in exponentiating the Lie
derivatives, leads to non--local commutation relations, which
reintroduce the longitudinal and temporal coordinates, and thus
are unacceptable. The completion of the Hamiltonian structure in
the general case is left for further studies.

This paper will be organized as follows: In Sect. \ref{S_SCATT} we
shall describe the scattering problem that we have in mind and its
formulation in the CGC formalism. This will also give us the
opportunity to introduce some notations and fix our conventions.
In Sect. \ref{S_RGE} we shall explain the general philosophy of
the RG approach underlying the effective theory for the CGC. The
discussion will be mostly qualitative, with the aim of
anticipating and thus clarifying the physical interpretation of
the subsequent, more technical, developments. The next two
sections will provide the mathematical formulation of the RG: In
Sect. \ref{S_SEFF} we shall derive general formulae for the
effective action encoding the correlations induced after one step
in the quantum evolution. The action is manifestly
gauge--invariant and is expressed in terms of the background field
propagator for the semi--fast gluons, that we shall construct
(within the approximations and for the physical regime of
interest) in Sect. \ref{S_prop}. The self--duality will appear
already in Sect. \ref{S_SEFF}, as a symmetry relating two
different, but equivalent, expressions for the effective action.
The first application of the general formalism will be given in
Sect. \ref{S_JIM}, where we shall provide a streamlined derivation
of the JIMWLK Hamiltonian. This derivation is considerably shorter
then the previous ones in the literature \cite{JKLW97,RGE}, thus
demonstrating the efficiency of the present formulation of the RG.
Then, in Sect. \ref{S_GEN} we turn to the general case and
construct the effective action as a functional of the two types of
Wilson lines. We shall derive four different, but equivalent,
versions for this action. The self--duality is now manifest, as
the invariance of the action under the exchange between
longitudinal and temporal Wilson lines. We shall also check that
the JIMWLK action is indeed obtained from the general action in
the limit where fluctuations become negligible. Finally, in Sect.
\ref{S_BREM} we consider the limiting case of a dilute regime and
thus deduce the Hamiltonian for bremsstrahlung. We construct the
associated Poisson brackets and explain how to use them in order
to derive evolution equations for observables like the scattering
amplitudes. Sect. 9 contains our conclusions.

\section{The scattering problem in the CGC formalism}
\setcounter{equation}{0} \label{S_SCATT}

Let us start with a description of the scattering problem that we have
in mind. We shall consider a high energy scattering in which the target
is a right--mover (it propagates in the positive $z$, or positive
$x^+$, direction), whereas the projectile is a left--mover (it
propagates towards negative $z$, or positive $x^-$). Note that,  for
the target, $x^+$ plays the role of time and $x^-$ that of the
longitudinal coordinate, while for the projectile the roles of $x^+$
and $x^-$ are interchanged. Furthermore, we shall view the scattering
in a Lorentz frame in which the target carries most of the total
energy, so that its wavefunction is highly evolved
--- it contains many small--$x$ gluons --- and can be described as a
color glass condensate (see below). In that frame, the projectile is a
relatively simple object, like a quark--antiquark pair in a colorless
state (a `color dipole'), or a collection of several such dipoles.

Since the gluon density in the target is typically high, the projectile
will undergo multiple scattering, which can be resummed in the eikonal
approximation: A quark (or antiquark) which propagates through the
target preserves a straightline trajectory but acquires a color
precession; that is, its wavefunction gets multiplied by a Wilson line
built with the projection of the color field of the target along the
trajectory of the quark. For a left--moving quark with transverse
coordinate $\x$, this Wilson line reads :
 \be \label{up}
  V^\dag(\x)={\mbox P}\exp \left\{
ig\int dx^- A^+_a(x^+\simeq 0,x^-,\x)\,t^a \right\}\,,
 \ee
where the $t^a$'s are the generators of the SU$(N_c)$ algebra in the
fundamental representation and the symbol P denotes the $x^-$ ordering
of the color matrices $A^+_a(x^-)t^a$ in the exponent, from right to
left in increasing order of their $x^-$ arguments. The integration runs
formally over all the values of $x^-$, but in reality it is restricted
to the longitudinal extent of the target, which is localized near
$x^-=0$ because of Lorentz contraction. By the same argument, the
projectile is localized near $x^+=0$, so it probes the target field
$A^+$ at small $x^+$, as also indicated on Eq.~(\ref{up}).

In fact, the field $A^+$ is slowly varying in $x^+$, since it is
generated by fast moving color sources --- the quarks and gluons inside
the target --- whose internal dynamics is slowed down by Lorentz time
dilation. Thus, the target field can be treated as constant over the
duration of the scattering. This motivates the {\em color glass}
picture to be developed below, in which the scattering observables are
first computed for a given configuration of the color fields in the
target, and then averaged over the latter.

Specifically, the $S$--matrix for the scattering between a color dipole
and the color field $A^+$ is obtained as the color average of the
product of two Wilson lines: one for the quark and the other one for
the antiquark (with transverse coordinates $\x$ and $\y$, respectively)
:
 \be\label{SdipoleA} S(\x,\y;A^+) \,=\,
\frac{1}{N_c}\,\tr (V^\dag_{\x} V_{\y}). \ee The physical $S$--matrix
for dipole--hadron scattering is finally obtained after averaging over
all the configurations of the color fields $A^+$ in the target. Within
the CGC effective theory, this average is computed as follows:
\be\label{Sdipole} \langle
S(\x,\y)\rangle_\tau\,=\,\frac{1}{N_c}\big\langle \tr (V^\dag_{\x}
V_{\y})\big
 \rangle_\tau\,\equiv\,\int D[\rho]\,\,
 {\cal W}_{\tau}[\rho]\,\frac{1}{N_c}\,\tr (V^\dag_{\x}
 V_{\y}). \ee
This formula makes it explicit that the field $A^+$ is created by
`color sources' (quantum fluctuations in the target wavefunction) with
large longitudinal momenta $k^+$, whose lifetimes are much larger than
the collision time.
In the computation of scattering amplitudes, these sources can be
effectively replaced by a classical color current $J^\mu_a=\delta^{\mu
+}\rho_a$, where the charge density $\rho_a\equiv\rho_a(x^-,\x)$ is
{\em static}, i.e., independent of $x^+$, but {\em random}, as it
corresponds to any of the possible configurations of the fast moving
partons. The correlations of $\rho$ are encoded in the {\em weight
function} ${\cal W}_{\tau}[\rho]$ --- a functional probability
distribution which depends upon the target rapidity\footnote{In the
considered frame, $\tau$ is essentially the same as the rapidity gap
between the target and the projectile.} $\tau \sim \ln s$ because of
{\em quantum evolution} :

The fast partons which represent the color sources have longitudinal
momenta within the range $P^+\gg k^+\gg \Lambda$, where $P^+$ is the
total momentum of the target and $\Lambda$ is the scale at which the
lifetime of a virtual excitation becomes of the order of the collision
time. When increasing the energy $s$ of the collision, $P^+$ increases
like $\sqrt{s}$, while $\Lambda$ remains constant (since this is fixed
by the properties of the projectile). Thus, the longitudinal
phase--space $P^+\gg k^+\gg \Lambda$ available for quantum evolution is
increased, leading to an enhanced gluon radiation, and thus to the
generation of new color sources. The effects of this evolution will be
discussed in detail and computed in the next sections.

In particular, in the dilute regime at not so high energies, the target
field is weak and Eq.~(\ref{SdipoleA}) can be evaluated by expanding
the exponentials in the Wilson lines. To lowest non--trivial order in
this expansion one obtains the scattering amplitude in the two--gluon
exchange (or single scattering) approximation:
 \be\label{Tdipole_weak}
  T(\x,\y)\equiv 1 - S(\x,\y) \, \simeq \,
  \frac{g^2}{4N_c}\,
 (A^+_a(\x)-A^+_a(\y))^2\,,\ee
where
 \be\label{AT} A^+_a(\x)\,\equiv\,\int
 dx^-\,A^+_a(x^+=0,x^-,\x)\ee
is the effective color field in the transverse plane, as obtained after
integrating over the longitudinal profile of the target. Note that the
ordering of the Wilson lines in $x^-$ plays no role in this
approximation, because of the symmetry of the color trace: ${\rm tr}
(t^a t^b)=\frac{1}{2}\,\delta^{ab}={\rm tr} (t^b t^a)$. However, this
ordering {\em does} matter in the strong field regime in which multiple
scattering is important and the complete equation (\ref{SdipoleA}) must
be used: successive collisions do not commute with each other (as they
involve color matrices), so they resolve the longitudinal structure of
the color field in the target. Thus in the construction of the
effective theory one should also keep trace of the correlations of the
color sources in $x^-$. But in the approximations of interest at high
energy, these correlations turn out to be relatively simple and can
always be embedded via Wilson lines. In particular, the effective
theory and its evolution can be fully written in terms of Wilson lines
\cite{B,W}, in which case the longitudinal coordinate $x^-$ is not
explicit anymore.

\section{Quantum evolution: The general philosophy}
\setcounter{equation}{0} \label{S_RGE}

In this section we shall explain the general structure of the quantum
evolution and describe our strategy towards performing the calculation.
This strategy relies on the {\em renormalization group} (RG) in QCD at
small--$x$, a technique which allows one to integrate out quantum
fluctuations in the evolution with increasing energy by exploiting the
separation of (longitudinal and temporal) scales specific to the
high--energy problem. This technique generalizes the BFKL evolution
\cite{BFKL} and has been originally developed in Refs.
\cite{JKLW97,RGE} in relation with the JIMWLK equation. In what follows
we shall further develop this technique and apply it to the
construction of more general evolution equations, from which the JIMWLK
equation will emerge as a special limit. The discussion in this section
is purely qualitative: we shall explain the structure of the quantum
corrections and emphasize some general properties which will facilitate
the physical interpretation of the calculations in the next sections.
We shall thus anticipate from physical considerations some structural
properties of the evolution equations that will be later confirmed by
explicit calculations.

As explained in the previous section, our objective is to
construct an effective theory for gluon correlations in the target
wavefunction at some soft longitudinal momentum scale $\Lambda$
with $\Lambda\ll P^+$. For instance, to compute scattering
amplitudes at the scale $\Lambda$, we need\footnote{More
precisely, we only need the correlations of the Wilson lines, like
Eq.~(\ref{Sdipole}), but for the purpose of describing the quantum
evolution it is more transparent to think in terms of color fields
and their sources.} $n$--point functions like $\lan A^+(x)
A^+(y)\cdots\ran_\Lambda$, where the subscript $\Lambda$ means
that the fields $A^+$ inside the brackets carry soft longitudinal
momenta $k^+\sim \Lambda$. At high energy, such {\em soft} fields
are predominantly generated by relatively {\em fast} color
sources, with longitudinal momenta $p^+$ such that $P^+\gg p^+\gg
\Lambda$. Indeed, due to the infrared singularity of the
bremsstrahlung spectrum $\propto dp^+/p^+$, the contribution of
these sources to the correlations at scale $\Lambda$ will be
enhanced by powers of the large energy logarithm $\ln
(P^+/\Lambda) \sim\ln s$, and thus will dominate over the genuine
quantum fluctuations with $k^+\sim \Lambda$. Therefore, in the
leading logarithmic approximation (LLA) to which we shall restrict
ourselves in what follows, the soft fields at scale $\Lambda$ are
entirely generated by fast color sources with $p^+\gg \Lambda$ and
inherit the correlations of the latter. To the same approximation,
the relevant `color sources' are gluons which are themselves
radiated by other gluons with even larger longitudinal momenta.
This separation of scales makes it natural to use an RG analysis
in which the fast gluon fluctuations are integrated out in layers
of $p^+$ and replaced by an effective color charge density
$\rho_a(x^-,\x)$ which generates the same correlations as the
quantum gluons at the soft scale $\Lambda$.

Before we describe the quantum evolution, let us briefly review the
solution to the classical equations of motion which rely the soft
fields to $\rho$ (this solution will be needed later). The relevant
equations are the Yang--Mills equations with a source
$J^\mu_a=\delta^{\mu +}\rho_a$ ,
  \be (D_{\nu} F^{\nu \mu})_a(x)\, =\, \delta^{\mu +}
 \rho_a(\vec x)\,, \label{YMJ} \ee
where $\vec x=(x^-,\x)$ and $D_{\nu} = \partial_\nu - ig A_\nu^a T^a$
with $(T^a)_{bc}= -if_{abc}$. Because of the relatively simple
structure of the current in the r.h.s., these equations can be solved
explicitly (at least in specific gauges). Namely, it is consistent with
Eq.~(\ref{YMJ}) to search for a solution having the following
properties \cite{CGCreviews} \be\label{YMprop} F^{ij}_a=0,\qquad
 A^-_a=0,\qquad
 A^+_a,\,A^i_a\,:\,\,{\rm independent\ of\ } x^+\,. \ee
Then the only non--trivial field strength  is 
$F^{+i}_a$. After also imposing a gauge condition, the classical
solution involves just one independent field degree of freedom. This
becomes manifest in the {\em Coulomb gauge} $\grad^i \tilde {A}^i =0$,
where Eq.~(\ref{YMJ}) reduces to
 \be\labe{EQTA} -
\nabla^2_\perp
 \widetilde A^+_a({\vec x})\,=\,{\widetilde \rho}_a(\vec x)\,,\ee
which is the light--cone analog of the Poisson equation. Note that we
use a tilde to denote the classical source and field in the Coulomb
gauge.  In fact, the Coulomb gauge potential will appear so often in
what follows that it becomes convenient to introduce a shorter notation
for it: $\widetilde A^+_a \equiv \alpha_a$. Eq.~(\ref{EQTA}) is
immediately solved as
 \be\label{alpha} \alpha_a (x^-, \x) =
 \int \frac{d^2\y}{4\pi}\, \ln\frac{1}{(\x -
 \y)^2\mu^2}\, \widetilde\rho_a  (x^-, \y)\,\equiv \int d^2\y\,
 \Delta(\x-\y)\,\widetilde\rho_a  (x^-, \y), \ee
where the infrared cutoff $\mu$ is necessary to invert the Laplacian
operator in two dimensions, but it will eventually disappear from the
physical results. The field strength in this gauge is obtained as
$\tilde {F}^{+i}_a\,=\,-\partial^i\widetilde \alpha_a$.

We shall also need later the classical solution in the light--cone (LC)
gauge $A^+_a=0$. This is constructed from the above solution in the
Coulomb gauge via a gauge rotation
 \be {A}^i(x^-,\x)\,=\,\frac{i}{g} \,
 U(x^-,\x)\,\partial^i U^{\dagger}(x^-,\x)\,,\label{tpg} \ee
 with
 \be
 U^{\dagger}(x^-,\x)=
 {\rm P} \exp
 \left \{ig \int_{-\infty}^{x^-} dz^-\,\alpha_a (z^-,\x) T^a
 \right \}.\label{UTA}
 \ee
 where we use matrix notations appropriate for the adjoint
representation (${A}^i={ A}^i_a T^a$, etc). Note the boundary
condition: $A^i_a(x^-,\x)\to \, 0$ for $x^-\to\,-\infty$.

Eqs.~(\ref{alpha}), (\ref{tpg}),  and (\ref{UTA}) together provide an
explicit expression for the  LC--gauge solution ${A}^i$ in terms of the
color source ${\tilde \rho}\,$ in the Coulomb gauge. Note that the
target average in equations like (\ref{Sdipole}) can be computed by
using the classical solution in any gauge, since the measure, the
weight function, and the physical observables are all gauge invariant.

\begin{figure}
\begin{center}
\centerline{\epsfig{file=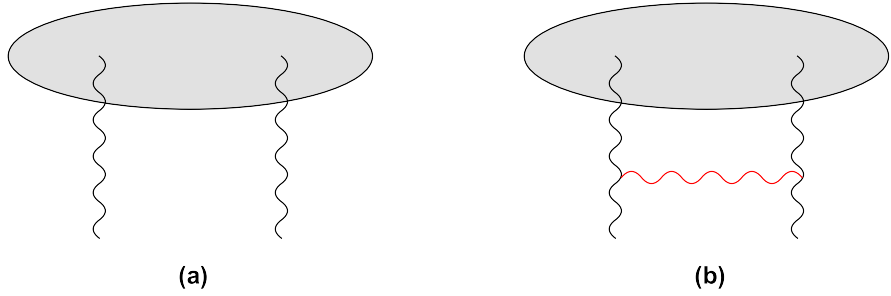,height=4.cm}} \caption{\sl The
2--point function in the effective theory (a) and its BFKL evolution in
one step (b). \label{BFKLfig}}
\end{center}
\end{figure}

We now return to the problem of the quantum evolution and start with
the simplest case, that of the evolution of the 2--point function
(\ref{Tdipole_weak}) in the dilute regime. In the LLA of interest, this
evolution is described by the BFKL equation \cite{BFKL}, and is
schematically illustrated in Fig. \ref{BFKLfig}.

Fig. \ref{BFKLfig}.a represents the  2--point function in the effective
theory at scale $\Lambda$ : the color fields $A^+$ are radiated from
the classical charge density $\rho$ and the average over $\rho$ (with
weight function ${\cal W}_{\Lambda}[\rho]$) is represented by the upper
blob. The one--step BFKL evolution of this 2--point function is
represented in Fig. \ref{BFKLfig}.a: this rung diagram is a typical
quantum correction which becomes important when we probe correlations
at the lower scale $b\Lambda$, with $b\ll 1$. These correlations have
two sources: the direct emission of a gluon with $k^+\sim b\Lambda$
from the classical source at scale $\Lambda$ (cf. Fig. \ref{BFKLfig}.a)
and the induced radiation from the {\em semi--fast} quantum gluons with
intermediate momenta $\Lambda \gg p^+ \gg b\Lambda$ (the rung in Fig.
\ref{BFKLfig}.b). The quantum process gives a correction of order
$\alpha_s\ln (1/b)$, and we shall assume that $\alpha_s\ln (1/b)\ll 1$
in order for perturbation theory to apply. The purpose of the
renormalization group analysis is to replace the two contributions in
Figs. \ref{BFKLfig}.a and b by a single classical contribution similar
to Fig. \ref{BFKLfig}.a, but with a modified weight function ${\cal
W}_{b\Lambda}[\rho]$. By computing the difference ${\cal W}_{b\Lambda}
- {\cal W}_{\Lambda} \sim \alpha_s\ln (1/b)$, one can establish the
{\em renormalization group equation} (RGE) whose solution permits to
iterate an arbitrary number of evolution steps. Introducing the
rapidity $\tau\equiv \ln (\Lambda/P^+)$, so that $d\tau = \ln (1/b)$
for one step of RG, it is clear that $\tau$ plays the role of the
evolution `time'.

It turns out that, quite generally, the RGE can be cast in Hamiltonian
form, and can be applied either to the weight function
 \be\label{RGEgen1}
 \frac{\partial}{\partial \tau} \,{\cal W}_\tau[\rho]= -
     H\Big[\rho, \frac{\delta}{i\delta \rho}\Big]
     {\cal W}_\tau[\rho] \, ,
     \ee
or directly to the interesting observables, for example the dipole
$S$--matrix     (\ref{Sdipole})
      \be
     \frac{\partial}{\partial \tau} \, \lan S[\rho]\ran_\tau= -
     \Big \lan H\Big[\rho, \frac{\delta}{i\delta \rho}
     \Big] \,S[\rho]\Big \ran_\tau\, ,
  \label{RGEgen2}
  \ee
where $H$ is a Hermitian functional differential operator that we shall
refer to as the  `Hamiltonian'. By inspection of the diagrams in Fig.
\ref{BFKLfig}, it should be already clear what is the general structure
of the BFKL Hamiltonian: When applied to a $n$--point function of the
charge density $\rho$, this should annihilate two $\rho$'s and replace
them by two new ones which interact with each other with the exchange
of a BFKL rung. This suggests that
 \be\label{HBFKL0}
  H_{\rm BFKL} \,=\,\frac{1}{2}\,\cal{K}_{\rm BFKL}\,
  \frac{\delta}{\delta \rho}\,\rho\,\rho\,
  \frac{\delta}{\delta \rho}\,,\ee
in schematic notations in which $\rho$ should be understood as the {\em
two--dimensional} color charge density, obtained as in Eq.~(\ref{AT})
 \be\label{rhoT}
 \rho_a(\x)\,\equiv\,\int
 dx^-\,\rho_a(x^-,\x).\ee
Indeed, there is no multiple scattering in the BFKL evolution, so the
longitudinal structure can be trivially integrated out leaving a
non--trivial dynamics in the transverse plane alone.

Note two important features of Eq.~(\ref{HBFKL0}) which are more
general than the BFKL approximation and will be recovered for the
general equations below. First, the fact that the r.h.s. of
Eq.~(\ref{HBFKL0}) is a total derivative with respect to $\rho$ is
necessary to ensure probability conservation: this guarantees that the
normalization condition $\int D[\rho]\, {\cal W}_{\tau}[\rho]=1$ is
preserved by the evolution. Second, note the specific ordering of the
operators $\rho$ and ${\delta}/{\delta \rho}$ (which do not commute
with each other) : this is obtained after subtle cancellations between
`real' and `virtual' corrections in the quantum calculation (see e.g.
Refs. \cite{W,RGE}), but is in fact imposed by gauge symmetry
\cite{ODDERON}. Thus, in order to obtain the evolution Hamiltonian, it
is in fact sufficient to compute the real correction (so like the real
gluon emission in Fig. \ref{BFKLfig}.b) and then order the operators in
such a way to ensure gauge symmetry. This is the strategy that we shall
adopt in what follows.

\begin{figure}
\begin{center}
\centerline{\epsfig{file=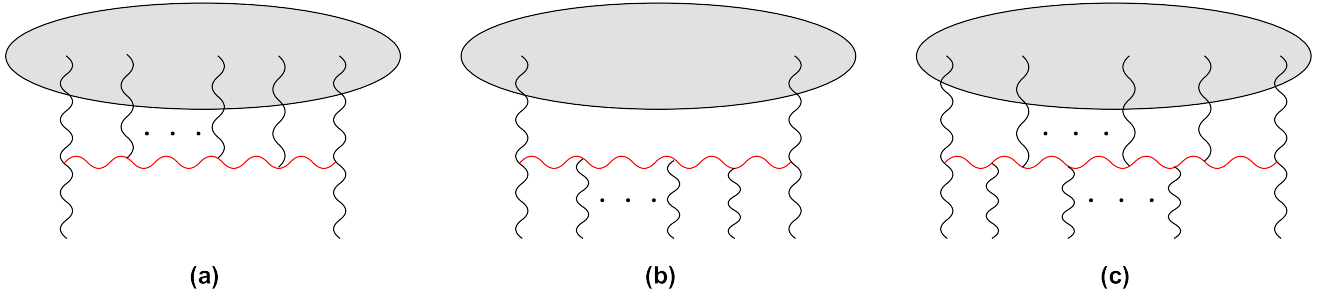,height=4.cm}} \caption{\sl One--step
quantum evolution with gluon number changing vertices: (a) a $n
\rightarrow 2$ merging process which contributes to the JIMWLK
Hamiltonian; (b) a $2 \rightarrow n$ splitting process as included in
the BREM Hamiltonian; (c)  a general $m\rightarrow n$ process.
\label{JBfig}}
\end{center}
\end{figure}

Let us turn now to the more interesting case where the target field is
strong  (this is the typical situation at high energy). This entails
non--linear effects both in the classical equations of motion
(\ref{YMJ}) and in the quantum evolution. Indeed, in this case, the
semi--fast gluon in Fig. \ref{BFKLfig}.b propagates in the background
of the strong classical field generated by the source $\rho$ at scale
$\Lambda$ and undergoes {\em multiple scattering} off this field.
Accordingly, the propagator of the semi--fast gluon must be computed to
all orders in the background field. A typical diagram included in this
resummation is shown in Fig. \ref{JBfig}.a. Clearly, this describes
{\em gluon recombination} : $n$ gluons (with $n\ge 2$) are merging into
two. Such merging processes are responsible for the saturation of the
gluon distribution and the formation of a color glass {\em condensate}.

In order to resum these processes to all orders, it is useful to notice
that the kinematical conditions are satisfied for the use of the {\em
eikonal approximation} . Indeed, as compared to the fast color sources
responsible for the background field, the `semi--fast gluons' are
relatively slow. In the rest frame of the background field (which is
the frame in which the eikonal approximation becomes most intuitive),
the semi--fast gluons propagate in the negative $z$, or positive $x^-$,
direction, so like the projectile in the scattering problem considered
in the previous section. Therefore, the gluon propagator, and also the
evolution Hamiltonian, will depend upon the background field via the
same Wilson lines as in Eq.~(\ref{up}), except that these are now
rewritten in the adjoint representation.

We thus see that, in the presence of non--linear effects, the quantum
evolution is able to discriminate the longitudinal structure of the
color source, so like the scattering with an external projectile.
This brings us to the issue of the $x^-$ correlations generated by the
quantum evolution. Because of the separation of scales in the problem,
these correlations turn out to be quite trivial: they merely show that
the color source $\rho_a(x^-,\x)$ extends towards larger values of
$x^-$ when increasing $\tau$ \cite{RGE}. Namely, by the uncertainty
principle, the color source generated after integrating out the fast
partons with momenta $p^+ \gg \Lambda$ is localized near $x^-=0$ within
a distance $\Delta x^- \sim 1/\Lambda$. With the boundary conditions
that we shall use in our calculation, and which are the same as in Ref.
\cite{RGE}, the support of the source lies fully at positive values of
$x^-$, namely at $0 \le x^- \le x^-_\tau$ with $x^-_\tau\sim 1/\Lambda
= (1/P^+) \rme^\tau$. When a new layer of semi--fast gluons is
integrated out, the additional color charge generated in this way is
localized at $1/\Lambda \simle x^- \simle 1/b\Lambda$, and thus has no
overlap with the previously existing color source. Thus, the quantum
evolution proceeds by adding new layers to the color source at larger
values of $x^-$, whereas the charge density in the inner layers
--- as generated in the previous steps --- is not modified. This means
that the correlations induced by the evolution are essentially local in
$x^-$ \cite{PATH}.

The previous discussion implies that, in a Wilson line like
Eq.~(\ref{up}), the support of the integration is truly restricted to
$0 \le x^- \le x^-_\tau$ and, moreover, the net effect of increasing
$\tau$ by $d\tau$ is an infinitesimal {\em gauge rotation} of the
Wilson line at its large--$x^-$ end
\begin{equation}\label{Utau1}
V^\dagger_{\tau}({\bm{x}})=\,{\rm P} {\rm
exp}\left\{ig\int_{0}^{x^-_\tau} dx^- \alpha(x^-,\bm{x})\right\}\quad
\longrightarrow\quad V^\dagger_{\tau+d\tau}({\bm{x}})=\,{\rm e}^{{\rm
i}g\int dx^-
\delta\alpha_\tau(x^-,\bm{x})}\,V^\dagger_{\tau}({\bm{x}})\,,
 \end{equation}
where $\delta\alpha^a_\tau$ has support in the outer layer at
$1/\Lambda \simle x^- \simle 1/b\Lambda$, but its detailed $x^-$
dependence is irrelevant in the approximations of interest.
Correspondingly, the functional derivatives in the RGE in the strong
field regime are taken with respect to the color source (or field) in
the outmost bin in $x^-$
  \be\label{HJIMWLK0}
  H_{\rm JIMWLK} \,=\,\frac{1}{2}\,
  \frac{\delta}{\delta \alpha_\tau}\,\chi_{\rm JIMWLK}[V, V^\dagger]\,
  \frac{\delta}{\delta \alpha_\tau}\,.\ee
This is the JIMWLK Hamiltonian whose first complete derivation
(including the subtle issue of the $\tau$--dependence of the functional
derivatives) has been given in Ref. \cite{RGE}. Mathematically, the
functional derivatives which appear in Eq.~(\ref{HJIMWLK0}) are really
{\em Lie derivatives} with respect to the Wilson lines $V$ and
$V^\dagger$ \cite{PATH}. This is explicit in the independent derivation
of the JIMWLK Hamiltonian by Weigert \cite{W} from the Balitsky
equations \cite{B}.

Thus, in this strong field regime, the evolution Hamiltonian is
naturally expressed in terms of Wilson lines and functional derivatives
with respect to the latter, and not directly in terms of the color
source $\rho_a(x^-,\bm{x})$. This is indeed the correct way to think
about the high--energy evolution: as a renormalization group flow in a
two--dimensional Hilbert space with a non--trivial geometry (namely,
the group manifold spanned by the Wilson lines). On the other hand,
given the separation of scales in the problem, one cannot address
questions about the detailed longitudinal structure of the color source
(like, e.g., computing correlations between the values of
$\rho_a(x^-,\bm{x})$ at different points $x^-$) : only the correlations
of the Wilson lines make sense. We shall recover this feature in more
general situations below.

Consider now the $2 \rightarrow n$ splitting diagram in Fig.
\ref{JBfig}.b, in which the $n$ emerging gluons are all soft (i.e.,
they carry momenta  $k^+\sim b\Lambda$). Physically, this diagram
describes {\em gluon bremsstrahlung} in the process of the BFKL
evolution. Through such processes, $n$--point correlation functions
with $n > 2$ get built from the 2--point function in the course of the
quantum evolution. Gluon splitting has not been included in the
analysis leading to the JIMWLK equation since formally this is a
higher--order effect (as compared to the direct emission of $n$ gluons
from the classical source $\rho$) in the strong field regime. However,
this argument implicitly assumes that the higher correlations for
$\rho$ have been already included in the weight function ${\cal
W}_{\tau}[\rho]$. But when starting the evolution in the dilute regime
(at low energy, or high transverse momenta), the  higher--point
correlations are originally absent, and they can only get built from
the 2--point function. Thus, gluon splitting is in fact the dominant
process for generating correlations in the dilute regime, and as such
it has also a strong influence on the dynamics in the high--density
regime, since the non--linear effects mix various correlations with
each other, as obvious in Fig. \ref{JBfig}.a.

Focusing on the dilute regime for the time being, we would like to
understand how to include bremsstrahlung in the BFKL evolution. To that
aim, we notice that a diagram like Fig. \ref{JBfig}.b can be
interpreted as describing the propagation of the semi--fast gluon in
the background of a soft color field (the radiated gluons). The
background field is physically weak, as it corresponds to quantum
radiation, but the associated effects will be resummed here to all
orders (that is, we shall formally treat this field as being strong)
since we want to generate all the $2 \rightarrow n$ processes for any
$n$ at once. Then the propagation of the semi--fast gluon can be again
treated in the eikonal approximation. The longitudinal momentum of this
gluon, comprised in the range $\Lambda \gg p^+ \gg b\Lambda$, is much
larger than that of the radiated fields, so in the rest frame of the
latter it will appear as a fast projectile moving in the positive $x^+$
direction. Accordingly, the semi--fast gluon couples to the minus
component $\delta A^-_a$ of the background field, via the {\em
temporal} Wilson lines
 \begin{align}\label{Wdef}
    W(\bm{x})=
    {\rm P} \exp\left\{i g \int
    \dif x^+ \delta\!A^-_a(x^+,\bm{x})T^a\right\}.
 \end{align}
We have chosen to denote this background field as $\delta A^-_a$ to
emphasize that it represents a quantum fluctuation at the soft scale
$b\Lambda$, and not a classical field radiated by $\rho$ (as previously
mentioned, the minus component vanishes for the classical solution; cf.
Eq.~(\ref{YMprop})). Alternatively, in the scattering problem, $\delta
A^-_a$ can be interpreted as the field radiated by the projectile, and
which couples to the semi--fast gluon from the target. In the
subsequent calculations, we shall generally use the first
interpretation of $\delta A^-_a$ (as a quantum field), but both points
of view will be useful in interpreting the final results.

By definition, $\dA^-_a$ has low $k^+$ and therefore large $k^-$ (since
this is a nearly on--shell gluon): $k^- \sim Q^2/2k^+$, with $Q^2$ a
typical transverse momentum scale in the problem. Therefore, the field
$\dA^-_a$ is quasi--independent of $x^-$ (more precisely, its
respective variation can be neglected over the relatively small
longitudinal extent $\Delta x^-$ of the semi--fast gluon to which it
couples) and is localized near $x^+=0$, within a small distance $\Delta
x^+\sim 1/k^-$. If $\dA^-_a$ is the field of the left--moving
projectile, the same properties follow from the Lorentz contraction and
the time dilation of the projectile. This explains why we have omitted
the $x^-$ variable in writing Eq.~(\ref{Wdef}).

\begin{figure}
\begin{center}
\centerline{\epsfig{file=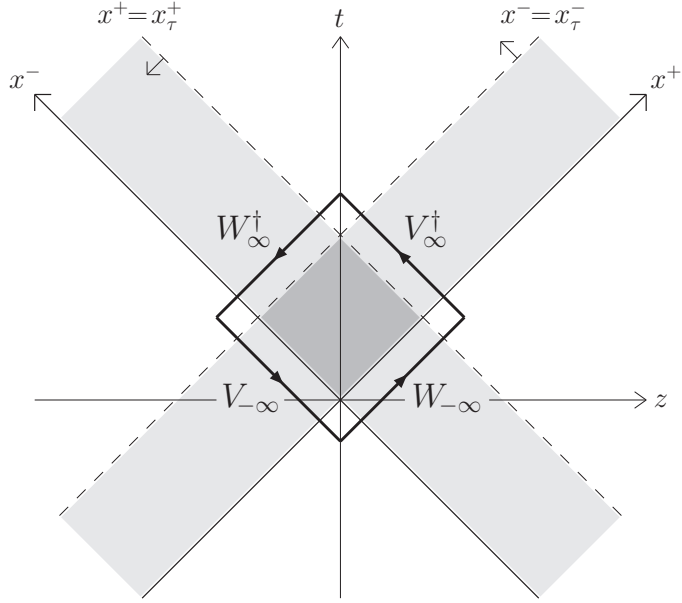,height=8.cm}}
\caption{\sl The distribution of the various background fields in
the $x^+ - x^-$ plane. The overlap region around the tip of the
light--cone is the interaction region relevant for quantum
evolution (see Sect. \ref{S_prop}). The contour surrounding the
interaction region will be explained in Sect. \ref{S_GEN}.
 \label{LCgeom}}
\end{center}
\end{figure}

The distribution of the various background fields in the longitudinal
direction and in time for the effective theory at rapidity $\tau$ is
illustrated in Fig. \ref{LCgeom}. The color source $\rho$ (and also the
Coulomb field $\alpha$ or the field strength $F^{+i}$) has support at
$0 \le x^- \le x^-_\tau$, with the upper limit $x^-_\tau$ increasing
with $\tau$. The soft field $\dA^-$ (and thus $F^{-i}$) is localized at
$0 \le x^+ \le x^+_\tau$, where $x^+_\tau$ decreases when increasing
$\tau$. Indeed the product $x^+_\tau x^-_\tau\sim 1/Q^2$ is constant
under evolution.

It should be clear by now that the result of resumming all splitting
diagrams like Fig. \ref{JBfig}.b is quadratic in $\rho$  and
non--linear to all orders in $\delta A^-$, upon which  it depends via
the Wilson lines $W$ and $W^\dagger$. It is furthermore intuitively
clear, and will be more formally justified in the next section, that
the effects of such diagrams on the evolution of the $n$--point
functions of the color charge $\rho$ can be obtained with the
replacement
 \begin{align}\label{atorhoBREM}
    \dA^-_a(x^+,\bm{x}) \to
    i\frac{\delta}{\delta \rho_a(x^+,\bm{x})}\,.
\end{align}
We thus anticipate that the evolution Hamiltonian describing
bremsstrahlung in the dilute regime should have the following general
structure:
 \be\label{BREM0}
  H_{\rm BREM} \,=\,\frac{1}{2}\,
  \rho\,\chi_{\rm BREM}[W, W^\dagger]\,
  \rho\,,\ee
where $\rho\equiv \rho_a(\x)$ is integrated over $x^-$, so like in
Eq.~(\ref{rhoT}), since in this regime there is no multiple scattering
in the $x^-$ direction. What is, however, more subtle now is the
$x^+$--dependence of $\rho$, which was irrelevant in the BFKL and the
JIMWLK approximations, but which becomes an issue in the presence of
multiple scattering in the $x^+$ direction: With the prescription
(\ref{atorhoBREM}), the Wilson line (\ref{Wdef}) generates functional
derivatives with respect to $\rho_a(x^+,\bm{x})$ at all the points
$x^+$ within the temporal extent $\Delta x^+$ of the fluctuating field
$\delta A^-_a$. Thus, the Hamiltonian (\ref{BREM0}) acts naturally on a
Hilbert space in which the color charge density carries also $x^+$
dependence. This shows that, in order to include correlations
associated with gluon splitting, one needs a generalization of the
original CGC effective theory which allows for time dependence. We
shall see later that such a generalization can be indeed given, but it
cannot account for the detailed temporal correlations that are {\em a
priori} encoded in a Hamiltonian like (\ref{BREM0}). This is a
consequence of the separation of scales in the problem: the semi--fast
gluon has small $k^-$ relative to the background field $\dA^-$, so it
cannot resolve the structure of the latter in $x^+$, but measures only
its overall distribution, via the temporal Wilson line $W$. This is
similar, but {\em dual}
--- in the sense of exchanging $x^-$ by $x^+$, $\wt A^+$ by $\dA^-$,
and thus $V^\dagger$ by $W$
--- to the JIMWLK problem, in which the semi--fast gluon is unable to
discriminate the detailed $x^-$--structure of the fast sources, but
measures them only via the longitudinal Wilson line $V^\dagger$.
Inspired by this analogy, one can already anticipate that, when
interpreted in the sense of RG, the BREM Hamiltonian (\ref{BREM0})
describes the evolution of the correlation functions for $W$ and
$W^\dagger$, while the external factors of $\rho$ act as Lie
derivatives on these Wilson lines. This is what we shall find indeed in
the subsequent analysis.

The previous discussion suggests an interesting duality between the
evolution in the dense and the dilute regimes, which in turn reflects a
mirror symmetry between the $n\rightarrow 2$ and $2 \rightarrow n$
vertices in Figs. \ref{JBfig}. a and b, respectively. Specifically, the
corresponding Hamiltonians should correspond to each other via the
following {\em duality transformation} :
 \be\label{DUAL}\hspace*{-0.5cm}
 x^-\longleftrightarrow \,x^+,\qquad
 \alpha^a(x^-,\bm{x})\,\longleftrightarrow \,
  i\frac{\delta}{\delta \rho_a(x^+,\bm{x})}\,,\qquad
 \frac{1}{i}\, \frac{\delta}{\delta \alpha^a(x^-,\bm{x})}\,
 \longleftrightarrow \,
 \rho_a(x^+,\bm{x})\,.\ee
Anticipated by earlier works on a  symmetric formalism of the
high--energy scattering problem in QCD \cite{VV93,LKSeff,LipatovS,B},
such a duality has been recently found to hold in the large--$N_c$
approximation \cite{BIIT05}, where the only non--trivial vertices are
the $1\rightarrow 2$ and $2 \rightarrow 1$ triple--pomeron vertices,
and the coordinates $x^+$ and $x^-$ play no dynamical role.
Furthermore,  Kovner and Lublinsky have argued  \cite{KL3} that the
duality between mergings and splittings should hold {\em exactly},
i.e., beyond the large--$N_c$ approximation, as a consequence of a more
general, {\em self--duality}, property of the general (unknown)
evolution Hamiltonian.

However, the authors of Ref. \cite{KL3} have mistaken $x^+$ for
$x^-$ in the bremsstrahlung problem, and thus they failed to also
recognize the $x^-\leftrightarrow x^+$ duality. But after
correcting for this misidentification, we shall see that the
self--duality idea is essentially right. The general evolution
Hamiltonian that we shall construct in the next sections turns out
to be invariant under a generalized duality transformation which
exchanges Wilson lines in $x^-$ (so like $V^\dagger$) with those
in $x^+$ (like $W$). Moreover, the expressions for $H_{\rm
JIMWLK}$ and $H_{\rm BREM}$ that we shall find are indeed dual
each other in the sense of Eq.~(\ref{DUAL}). Our explicit
construction will further clarify the origin and the physical
meaning of this duality.

The general Hamiltonian alluded to above includes all the $m \to
n$ processes for arbitrary integers $m\ge 2$ and $n\ge 2$ (see
Fig. \ref{JBfig}.c). To resum such processes one needs the
propagator of the semi--fast gluon in the presence of both types
of background field: the field $\wt A^+_a\equiv \alpha_a$
generated by the fast partons ($p^+\gg \Lambda$) and the field
$\delta A^-_a$ associated with the soft quantum fluctuations
($k^+\sim b\Lambda$). This problem is complicated because the
semi--fast gluons are at the same time slow relative to $\wt
A^+_a$ and fast  as compared to $\delta A^-_a$, so in general the
eikonal approximation cannot be applied. However, for the purpose
of RG, the propagator is needed only for points $x^+$ and $x^-$
near the origin of the light--cone, that is, in the region where
the fields $\wt A^+_a$ and $\delta A^-_a$ overlap with each other
(see Fig. \ref{LCgeom}). In this region, we shall be able to
construct the propagator, and then the evolution Hamiltonian. The
latter will be expressed in terms of Wilson lines (line--ordered
phases in both $x^-$ and $x^+$) which run around the interaction
region in Fig. \ref{LCgeom} and represent the basic degrees of
freedom. The general Hamiltonian turns out to be local in the
transverse coordinates. The non--localities in the standard
expressions for the JIMWLK \cite{RGE,W} and BREM \cite{KL3}
Hamiltonians (which here will be generated by expanding the
general Hamiltonian in appropriate limits) are merely related to
the non--locality of the relation (\ref{alpha}) between the color
field in the Coulomb gauge and its source.

\section{Quantum evolution: The effective action}
\setcounter{equation}{0} \label{S_SEFF}

With this section, we start our explicit construction of the evolution
Hamiltonian for QCD at high energy. This Hamiltonian must encompass
processes like that in Fig. \ref{JBfig}.c to all orders in the two
types of background field : the ones above and those below the
horizontal gluon line. The resummation of these diagrams can be
formulated as a one--loop calculation in the quantum version of the
effective theory for the CGC \cite{JKLW97,RGE}. This is a theory in
which the fast partons with longitudinal momenta $p^+\gg \Lambda$ are
represented by a random color charge density $\rho_a(x^-,\bm{x})$ with
weight function ${\cal W}_{\Lambda}[\rho]$ while the slow gluons with
momenta $\le\Lambda$ are still explicitly present as quantum
fluctuations. Thus, this effective theory allows for the calculation of
gluon correlations at any scale $k^+\le \Lambda$.

For instance, the 2--point correlation function is obtained as:
\begin{align}\label{AA}
    \langle {\rm T} A^{\mu}(x) A^{\nu}(y) \rangle =
    \int D [\rho] \,{\cal W}_{\Lambda}[\rho]\,
    \frac{\int^{\Lambda} D\delta\! A\, A^{\mu}(x) A^{\nu}(y)\,
     \exp(i S[A,\rho])}
    {\int^{\Lambda} D\delta\! A\,\exp(i S[A,\rho])}
\end{align}
where the symbol T denotes time--ordering and it is understood that the
two fields inside the brackets `live' both at the same soft scale,
i.e., $k^+_1\sim k^+_2 \le \Lambda$. Furthermore, the action
$S[A,\rho]$ describes the gluon dynamics in the presence of the
`external' (or `background') color source $\rho$, and the total field
$A^{\mu}$ is the sum of the background field generated by $\rho$ and
the quantum fluctuations $\delta\!A^{\mu}$ with momenta $p^+\le
\Lambda$. As we shall see, the background field itself is influenced by
the fluctuations. The action $S[A,\rho]$ is such that it generates the
classical field equations (\ref{YMJ}) in the limit where
$\delta\!A^{\mu}=0$, but its precise form will not be needed below.

The one--step quantum evolution consists in integrating out the
semi--fast fluctuations with momenta $\Lambda \gg | p^+ | \gg b\Lambda$
where $b\ll 1$ but such that $\alpha_s\ln (1/b)\ll 1$. To that aim, we
shall assume that the external fields in Eq.~(\ref{AA}) carry momenta
$k^+\sim b\Lambda$ and we shall compute the correlations induced at the
soft scale $b\Lambda$ after integrating out the semi--fast fields
$a^{\mu}$. That is, the total field within the action is now decomposed
as \be
    A^{\mu} = B^{\mu} + a^{\mu} + \delta\!A^{\mu},
 \ee
where $B^{\mu}$ is the background field generated by $\rho$,
$a^{\mu}$ represents the semi--fast fluctuations, and
$\delta\!A^{\mu}$ stands for the remaining fluctuations with
$k^+\le b\Lambda$. Note that the total field at the scale
$b\Lambda$ (so like the external field in Eq.~(\ref{AA})) is
simply $A^{\mu} = B^{\mu} + \delta\!A^{\mu}$.

After separating the soft from the semi--fast modes in the functional
integral, the quantum average in Eq.~(\ref{AA}) becomes:
\begin{align}\label{AA2}
    \int^{b \Lambda} D \delta \! A\, A^{\mu}(x) A^{\nu}(y)
    \int_{b \Lambda}^{\Lambda} D a \exp(i S[A,\rho]),
\end{align}
where the last integration in Eq.~(\ref{AA2}) defines the {\em
effective action} that we want to compute:
\begin{align}\label{Seff}
    \exp(i S_{\rm eff}) \equiv \int_{b \Lambda}^{\Lambda} Da\,
    \exp(i S[A,\rho]).
 \end{align}
This is an action for the soft fields $\delta\!A^{\mu}$ which depends
also upon $\rho$. From this action, the evolution Hamiltonian will be
eventually obtained via the replacement (\ref{atorhoBREM}) (or, rather,
a generalization of it), that we shall shortly justify.

So far, we have not specified the gauge in which the quantum theory is
defined. With the separation of scales in $k^+$ being not
gauge--invariant, this issue is indeed important. The partonic
interpretation of gluons is meaningful in the light--cone quantization
which is defined in the LC--gauge $A^+=0$, so this is the gauge that we
shall adopt in what follows. Note that the separation of scales in
$k^+$ is invariant under the residual gauge transformations in the
LC--gauge, which are independent of $x^-$. In fact, the separation of
scales can be also formulated in a gauge--invariant way, as we shall
later explain, but the use of the LC--gauge in the quantum calculation
turns out to be really convenient. We should stress however that our
calculation will be manifestly covariant with respect to the
`background' fields $B^{\mu}$ and $\delta\!A^{\mu}$, so the final
result for the effective action will be gauge--invariant.

In the LC--gauge and for $\delta\!A^{\mu}=0$, the classical solution
has only transverse components $A^i$ which are given by
Eq.~(\ref{tpg}); these are time--independent and satisfy $F^{ij}=0$.
But in order to compute diagrams like that in Fig. \ref{JBfig}.c we
need to preserve an explicit background field $\delta\!A^-$ (the
component of the soft field $\delta\!A^{\mu}$ which couples to the
semi--fast gluon), and this will also modify the classical field
problem, by introducing time--dependence. Namely, in the presence of
$\delta\!A^-$, the color source $\rho$ will be subjected to {\em color
precession} :
 \be\label{Jplus}
 \rho(\vec{x})\equiv \rho_a(\vec{x})T^a \ \longrightarrow \
 J^{+}(x^{+},\vec{x}) =
W_{x^{+},-\infty} \,\rho(\vec{x}) \,W_{x^+,-\infty}^{\dagger}\,,
 \ee
meaning that the fast partons represented by $\rho$ undergo eikonal
scattering off the field $\delta\!A^-$. In the equation above,
 \begin{align}\label{WilsonW}
    W_{x^{+},-\infty}(\bm{x})=
    {\rm P} \exp\left\{{i g \int_{-\infty}^{x^+}
    \dif z^+ \delta\!A^-(z^+,\bm{x})}\right\},
\end{align}
is the temporal Wilson line from $z^+ \to -\infty$ up to the actual
time $x^+$. Note that the current $J^\mu=\delta^{\mu +}J^{+}$ is
covariantly conserved, as it should:
 \be
 D_{\mu}J^\mu = D^- J^{+} =(\del^- -ig\delta A^{-})J^{+}=0\,.\ee
We have anticipated here that $B^-=0$, so that $A^{-}=\delta A^{-}$
indeed. From Eq.~(\ref{Jplus}), one sees that one can identify $\rho$
with the asymptotic value of the current $J^{+}$ at $x^+ \to -\infty$,
and in what follows it will be useful to write:
 \be\label{rhoinfdef}
 \rho_{-\infty}(\vec{x})\,\equiv \, J^{+}(x^{+}\to -\infty
 ,\vec{x}) = \rho(\vec{x}),\qquad
 \rho_{\infty}(\vec{x}) \,\equiv \, J^{+}(x^{+}\to \infty
 ,\vec{x})
 .\ee

The background field $B^\mu=\delta^{\mu i}B^i$ is defined as the
solution to the Poisson equation with the current (\ref{Jplus}) :
\begin{align}\label{saddle}
    \frac{\delta S}{\delta A^{-}}\Big|_{B^i + \delta\!A^{-}}=\, 0
    \quad\Longrightarrow\quad D_{\nu} F^{\nu +} = J^{+} =
    W_{x^{+},-\infty} \rho_{-\infty}(\vec{x}) W_{x^+,-\infty}^{\dagger},
\end{align}
where, with a slight abuse of notations, the background field
$A^\mu=\delta^{\mu i}B^i+ \delta^{\mu -}\delta\!A^-$ has been denoted
as $B^i + \delta\!A^{-}$.

A crucial observation for what follows is that
 \be \label{FPM}
 F^{-+} = -\del^+
 \delta\!A^- \simeq 0,\ee
which follows from the fact that the field $\delta\!A^-$ contains only
modes with very small longitudinal momenta and therefore it varies very
slowly with $x^-$. More generally, the statement that $F^{-+}=0$
represents the gauge--invariant expression of the separation of scales
in the problem. With this property, Eq.~(\ref{saddle}) simplifies to
\begin{align}\label{saddle2}
    D_i F^{i+} = - D_i \del^+ B^i = J^+,
\end{align}
which will be shortly used to determine the field $B^i$ as a
function of $\delta\!A^-$ and $\rho$. Notice the time dependence
in the above equation : this is {\em local} in $x^+$, so $B^i$ and
$J^{+}$ have the same time dependence, which in turn comes from
the color precession in Eq.~(\ref{Jplus}). Since the soft gauge
field $\delta\!A^-$ has support at small $x^+$ (cf. Fig.
\ref{LCgeom}), the temporal Wilson line (\ref{WilsonW}) looks like
a smooth $\Theta$--function in $x^+$ --- its variation with $x^+$
is concentrated in a small region above $x^+=0$. A similar
$x^+$--dependence then holds for both $J^{+}$ and $B^i$.

For the purpose of the one--loop calculation, the action $S$ must be
expanded to second order in the semi--fast fields $a^\mu$. We obtain,
with condensed notations,
\begin{align}\label{Sexpand}
    S\,=\,S[B^i + \delta\!A^\mu, \rho] +\,
      \frac{\delta S}{\delta A^i}\Big|_{B^i + \delta\!A^-} a^i +\,
      \frac{1}{2}\,a^{\mu} \,G_{\mu \nu}^{-1}[B^i + \delta\!A^-, \rho]
      \,a^{\nu}\,,
\end{align}
where
\begin{align}\label{dsdA}
    \frac{\delta S}{\delta A^i}\Big|_{B^i + \delta\!A^-}=
    D_{\nu}F^{\nu i}= D_{j}F^{j i} + D_{-}F^{- i} + D_{+}F^{+ i},
\end{align}
and
 \be\label{Ginv}
 G_{\mu \nu}^{-1,\, ab}(x,y)\,[B^i + \delta\!A^-, \rho]\,\equiv\,
 \frac{\delta^2 S}{\delta A_a^\mu(x) \delta A_b^\mu(y)}
 \Big|_{B^i + \delta\!A^-}\,.\ee
As indicated in the equations above, only the $\delta\!A^-$ component
of the soft field $\delta\!A^\mu$ needs to be kept in the terms
describing the coupling with $a^\mu$.

Using again the fact that $\del^+ \delta\!A^- \simeq 0$, one can easily
show that
\begin{align}\label{FF}
    D_{-}F^{- i} \simeq   D_{+}F^{+ i},
\end{align}
a relation which lies at the origin of the duality property, as we
shall later see.

Consider now the Poisson equation (\ref{saddle2}) which determines
$B^i$. As compared to the corresponding equation for $\delta\!A^-=0$
(the component $\mu=+$ of Eq.~(\ref{YMJ})), the present equation
involves time dependence via $J^+$. However, this dependence is local,
so the solution to Eq.~(\ref{saddle2}) is again of the form $B^i=(i/g)
U \del^i U^{\dagger}$ (cf. Eq.~(\ref{tpg})), except that now the Wilson
lines depend also upon $x^+$ :
\begin{align}\label{WilsonU}
    U^{\dagger}(x)= {\rm P} \exp \left\{ i g \int_{-\infty}^{x^-}
    \dif z^- \alpha(x^+,z^-,\bm{x})\right\}.
\end{align}
Here, $\alpha\equiv \tilde B^+$ is the background field in the Coulomb
gauge, defined by the time--dependent generalization of
Eq.~(\ref{EQTA}) :
  \be\label{WTA} -
\nabla^2_\perp \alpha(x)
 \,=\,{\widetilde \rho}(x)\,\equiv\,U^{\dagger}(x)\,J^+(x)\,U(x)
 \,,\ee
which is local in both $x^-$ and $x^+$. The field $B^i$ constructed
above is a two--dimensional pure gauge ($F^{ij}=0$), so
Eq.~(\ref{dsdA}) simplifies to
\begin{align}\label{dsdA2}
    \frac{\delta S}{\delta A^i}\Big|_{B^i + \delta\!A^-} \simeq
    2 D_{+}F^{+ i} \simeq 2 D_{-}F^{- i}.
\end{align}
Performing the Gaussian integration over the semifast modes $a$ in
Eq.~(\ref{Seff}), we finally arrive at the following expression for the
effective action: $S_{\rm eff}=S + \Delta S_{\rm eff}$ with
\begin{align}\label{Seff2}
   \Delta S_{\rm eff}[\delta\!A^-,\rho]\, =\, -\frac{1}{2} \int\limits_{x y}\,
    (2 D^- F^{+ i})_x^a \,G^{i j}_{ab}(x,y) \,(2 D^- F^{+ i})_y^b\,,
\end{align}
or, equivalently (within the present approximations),
\begin{align}\label{Seff21}
   \Delta S_{\rm eff}[\delta\!A^-,\rho]\, =\, -\frac{1}{2} \int\limits_{x y}\,
    (2 D^+ F^{- i})_x^a \,G^{i j}_{ab}(x,y) \,(2 D^+ F^{- i})_y^b\,,
\end{align}
In these equations, $iG^{i j}$ are the transverse components of the
time--ordered propagator of the semi--fast gluons:
 \be
 iG^{\mu \nu}_{ab}(x,y)\,[B^i + \delta\!A^-, \rho]\,\equiv\, \lan{\rm T}
 \,a^{\mu}_a(x)\,a^{\nu}_b(y)\ran,\ee
where the brackets denote the average over the quantum fields $a^{\mu}$
for fixed values of the background fields $B^i$ and $\delta\!A^-$. This
propagator can be obtained by inverting the differential operator in
the r.h.s. of Eq.~(\ref{Ginv}) in the LC gauge and with appropriate
(Feynman) boundary conditions.

Eqs.~(\ref{Seff2}) and (\ref{Seff21}) are the fundamental formulae
which will allow us to construct the evolution Hamiltonian in the next
sections. As anticipated, these expressions are manifestly invariant
under the gauge transformations of the background fields (with the
propagator $G^{i j}$ being covariant under such transformations). In
what follows, we shall use this symmetry to replace the LC--gauge
background fields $A^\mu=\delta^{\mu i}B^i+ \delta^{\mu -}\delta\!A^-$
by the Coulomb gauge ones $\wt A^\mu=\delta^{\mu +}\alpha + \delta^{\mu
-}\wt{\delta\!A}^-$, which are most directly related to the various
Wilson lines. In the last equation, $\widetilde{\delta\!A}^-$ is the
soft field in the Coulomb gauge:
\begin{align}\label{datilde}
    \wt{\dA}^-(x)= U^{\dagger}(x)
    \left[ \dA^-(x^+,\bm{x}) + \frac{i}{g} \del^- \right] U(x),
\end{align}
and depends on $x^-$ (unlike its LC--gauge counterpart) via the
corresponding dependence of the gauge rotation. That is,
$\widetilde{\delta\!A}^-$ looks like a smooth $\Theta$--function
in $x^-$. One can check on Eq.~(\ref{datilde}) that the condition
$\wt F^{-+} =0$ (cf. Eq.~(\ref{FPM})) remains satisfied, as it
should.

In fact, with the background fields in the Coulomb gauge, one has:
 \be
 \wt D^+ = \del^+ -ig\alpha,\quad \wt F^{+ i}=-\del^i \alpha
 ,\quad \wt D^- = \del^- -ig\wt{\dA}^-,
 \quad \wt F^{- i}=-\del^i\wt{\dA}^-\,.
 \ee
We shall later argue that, under the present approximations, the
transverse propagator
 \be\label{Grot}
 \wt G^{i j}_{ab}(x,y) [\,\alpha,\wt{\delta\!A}^-]\,=\,
 U^{\dagger}(x)\,\,G^{i j}_{ab}(x,y)\,[\,B^i, \delta\!A^-]\,\,U(y)
 \,,\ee
is invariant under the  exchange of plus and minus components for
coordinates and fields:
 \be\label{DUAL1}
 x^-\longleftrightarrow \,x^+,\qquad
 \alpha_a(x)\,\longleftrightarrow \,\wt{\delta\!A}^-_a(x).\ee
Therefore,  under the transformations (\ref{DUAL1}) the two expressions
for the effective action, Eqs.~(\ref{Seff2}) and (\ref{Seff21}), get
interchanged. But since these expressions are equivalent with each
other,  we conclude that the effective action is {\em invariant} under
the transformations (\ref{DUAL1}). This symmetry is the most general
form of the property that in the previous sections has been referred to
as {\em `self--duality'}.

The physical meaning of this symmetry is easy to understand. The same
one--loop calculation resumming the diagrams illustrated in Fig.
\ref{JBfig}.c can  be interpreted in two ways: \texttt{(i)} As the
evolution of the color fields $\wt A^+\equiv \alpha$ of the target with
decreasing $k^+$ and in the presence of the fields $\wt{\delta\!A}^-$
of the projectile (this is the interpretation {\em a priori} privileged
by our previous manipulations, and it corresponds to the expression
(\ref{Seff2}) for the effective action), or \texttt{(ii)} as the
evolution of the color fields $\wt{\delta\!A}^-$ of the projectile with
decreasing $k^-$ and in the presence of the fields $\wt A^+\equiv
\alpha$ of the target (this interpretation corresponds to
Eq.~(\ref{Seff21}) and amounts to reading a diagram like Fig.
\ref{JBfig}.c upside down). By virtue of {\em boost invariance}, the
two directions of evolution must be equivalent: the rapidity increment
$d\tau$ can be given to either the target or the projectile, with
identical results for the evolution of the scattering amplitudes, which
are  Lorentz--invariant. This equivalence is indeed manifest at the
level of our results, as the invariance of the effective action under
the change (\ref{DUAL1}) of the direction of evolution.

To conclude this section, let us justify the replacement
(\ref{atorhoBREM}) which will allow us to transform the effective
action (a functional of $\rho$ and ${\delta\!A}^-$) into an evolution
Hamiltonian involving $\rho$ and $\delta/\delta\rho$. This step is
essential for promoting the above one--loop calculation into a
systematic renormalization group (RG) procedure. The purpose of RG is
to shift the newly  induced correlations at the soft scale $b\Lambda$
from the effective action $\Delta S_{\rm eff}$ to the weight function
${\cal W}_{\tau}[\rho]$ for $\rho$. To that aim, it is preferable to
work in the Coulomb background gauge, since scattering observables are
directly related to $\wt A^+\equiv\alpha$, and not to $B^i$. Consider
then the evolution
 \be
 \langle {\rm T} \alpha(x) \alpha(y) \rangle_{b\Lambda}
  - \langle {\rm T} \alpha(x) \alpha(y) \rangle_{\Lambda}\ee
of the 2--point function when decreasing $k^+$ from $\Lambda$ to
$b\Lambda$. We would like to associate this evolution with a change
$\alpha\to \alpha+\delta\alpha $ in the classical field ; through
Eq.~(\ref{WTA}), this corresponds to a change $\wt\rho\to \wt\rho+
\delta\wt{\rho}$ in the charge density in the Coulomb gauge.
Physically, $\delta\wt{\rho}$ represents the color charge density of
the semi--fast gluons. To reproduce the correlations encoded into
$\Delta S_{\rm eff}$, the induced source $\delta\wt{\rho}$ must be a
random variable (for a fixed distribution $\wt\rho$ of the color charge
at scale $\Lambda$), with correlations:
 \be\label{rhoG}\hspace*{-0.6cm}
 \langle {\rm T} \delta\wt{\rho}(x_1)\delta\wt{\rho}(x_2)\cdots
 \delta\wt{\rho}(x_n) \rangle_{\wt\rho}\,=\,i^{n+1}
 \Gamma_n(x_1,x_2,\dots,x_n)
 \,\equiv\,
 \frac{i^{n+1}\,\delta^n \Delta S_{\rm eff}}{\delta \wt A^-(x_1)
 \delta \wt A^-(x_2) ...\delta \wt A^-(x_n)}\Big|_{\wt A^-=0},\nn
 \ee
where for more clarity we have temporarily denoted $\wt{\delta\!A}^-$
simply as $\wt{A}^-$. Indeed, the quantum calculation yields (for the
example of the 2--point function, once again)
 \be
 \langle {\rm T} \,\wt{\delta \! A}^+\!(x)\,
 \wt{\delta \! A}^+\!(y) \rangle_{\wt\rho}
 = \int^{b \Lambda} D \wt{\delta \! A}\,\,\,\wt{\delta \! A}^+\!(x)
 \,\wt{\delta \! A}^+\!(y)\,
     \exp\big\{i\big(S[\,\wt{\delta \! A}\,]+ \Delta S_{\rm eff}
     [\,\wt{\delta \! A}^-,\wt\rho\,]\big)\big\},
     \ee
where the path--integral in the r.h.s. is now evaluated in the Coulomb
gauge. To the accuracy of interest, this has to be computed in the
saddle point approximation and to order $\alpha_s\ln (1/b)$ (recall
that the induced action $\Delta S_{\rm eff}$ is itself of
$\cal{O}(\alpha_s\ln (1/b))$). This yields:
 \be
 \langle {\rm T} \,\wt{\delta \! A}^+\!(x)\,
 \wt{\delta \! A}^+\!(y) \rangle_{\wt\rho}
  \,\simeq \,i\int\limits_{u,v} iG^{+-}(x,u)\,iG^{+-}(y,v)\,\,
  \Gamma_2(u,v)\,,
 \ee
which should be compared to the corresponding correlation induced, via
the classical field equation (\ref{WTA}), by a change $\delta\wt{\rho}$
in the color source:
  \be
 \langle {\rm T} \delta\alpha(x) \delta\alpha(y) \rangle_{\wt\rho}
 \,= \,\int\limits_{u,v} \Delta(x-u)\Delta(y-v)\,\,
 \langle {\rm T} \,\delta\wt{\rho}(u)\,
 \delta\wt{\rho}(v) \rangle_{\wt\rho}\,.\ee
In the static limit $p^-=0$, $G^{+-}(x,u)=\Delta(x-u)$, and the above
correlations coincide with each other provided Eq.~(\ref{rhoG}) is
satisfied.

Finally, it is clear that the Hamiltonian which, when acting on
correlations of $\wt\rho$ (cf. Eq.~(\ref{RGEgen2})), generates the
correlations in Eq.~(\ref{rhoG}) is of the form (with $\dif\tau=\ln
1/b$) :
 \be\label{Hgen}
     -\dif\tau \,H\Big[\wt\rho,\, i\frac{\delta}{\delta \wt\rho}\Big]&\,=\,&
  \sum_n \frac{1}{n!}\,\int\limits_{x_i}\,\Gamma_n(x_1,x_2,\dots,x_n)
 \,\frac{i^{n+1}\,\delta^n }{\delta \wt\rho(x_1)
 \delta \wt\rho(x_2) ...\delta \wt\rho(x_n)}\nn
 &\,=\,& i\Delta S_{\rm eff}
     \Big[\,\wt\rho,\,\,\wt{\delta \! A}^-\!=\,
     i\frac{\delta}{\delta \wt\rho}\,\Big]\,.
 \ee
Thus, the evolution Hamiltonian is obtained from the effective action
via the replacement
 \begin{align}\label{tilatorho}
    \wt{\dA}^-(x) \ \longrightarrow\
    {i}\,\frac{\delta}{\delta \wt{\rho}(x)},
\end{align}
which generalizes the previous prescription (\ref{atorhoBREM}) by
including the dependence upon $x^-$ and by specifying that in the
general, strong field, regime\footnote{Recall that
Eq.~(\ref{atorhoBREM}) has been written for the dilute regime, in which
the distinction between the LC--gauge and the Coulomb gauge for the
background fields becomes irrelevant, since one can approximate
$U\approx U^\dagger\approx 1$ in any gauge rotation like
Eq.~(\ref{datilde}).}, this replacement works at the level of Coulomb
gauge quantities.

\section{The background field propagator}\label{S_prop}
\setcounter{equation}{0}

In this section we shall construct the background field propagator
which is needed to compute the expressions (\ref{Seff2}) or
(\ref{Seff21}) for the effective action. This is the propagator
$G^{ij}(x,y)$ for a semi--fast, transverse, gluon in the quantum
LC--gauge ($a^+=0$) and in the presence of a background field which can
be taken either in the LC--gauge ($A^\mu=\delta^{\mu i}B^i+ \delta^{\mu
-}\delta\!A^-$), or in the Coulomb gauge ($\wt A^\mu=\delta^{\mu
+}\alpha + \delta^{\mu -}\wt{\delta\!A}^-$), and which involves two
independent field degrees of freedom. The general expression of this
propagator is complicated and we shall not attempt to derive it here.
But for the present purposes, a crucial simplification comes from the
fact that Eqs.~(\ref{Seff2})--(\ref{Seff21}) involve the propagator
$G^{ij}(x,y)$ only at points $x$ and $y$ within the {\em interaction
region} in Fig. \ref{LCgeom}. By the `interaction region' we mean the
diamond--shape area around the tip of the light--cone where the
supports of $\rho$ and $\delta\!A^-$ (or, equivalently, of $\alpha$ and
$\wt{\delta\!A}^-$) overlap with each other. The essential property of
this region is that its extent both in longitudinal and in temporal
directions is very small as compared to the respective resolution
scales of the semi--fast gluon. Because of that, the propagator
$G^{ij}(x,y)$ is local in the transverse coordinates and has a
relatively simple structure in the other coordinates, as we shall
explain below.

Let us first establish the support properties of the integrands in
Eqs.~(\ref{Seff2}) or (\ref{Seff21}). Consider Eq.~(\ref{Seff2}), for
definiteness, and look at the vertices there: the field $F^{+ i}$ has
the same support as the color source $\rho$ at scale $\Lambda$, so it
is localized in $x^-$ within the region $0\simle x^- \simle 1/\Lambda$.
Furthermore, its (covariant) derivative $D^-F^{+
i}=(\del^--ig\delta\!A^-)F^{+ i}$ involves two terms, which are both
localized at small $x^+$, within the support of $\delta\!A^-$. This is
obvious for the second term, which is explicitly proportional to
$\delta\!A^-$, but it is also true for the first term $\del^- F^{+ i}$,
since the time--dependence of $F^{+ i}$ comes entirely from its
interaction with $\delta\!A^-$, as discussed after Eq.~(\ref{saddle2}).
Recalling that $\delta\!A^-$ has small $k^+\simle b\Lambda$ and thus
large $k^-\simge Q^2/2b\Lambda$ (with $Q^2$ a typical transverse
momentum scale), we deduce that $D^-F^{+ i}$ has support at $0\simle
x^+ \simle b\Lambda/Q^2$. To summarize:
 \be
 D^-F^{+ i}\ne 0\qquad{\rm for}\quad 0\simle x^- \simle
 1/\Lambda\quad{\rm and}\quad 0\simle x^+ \simle b\Lambda/Q^2.\ee
This is also the support of the vertex $D^+F^{- i}$ in
Eq.~(\ref{Seff21}). On the other hand, the semi--fast gluon has
$\Lambda \gg k^+\gg b\Lambda$, and thus typical resolution scales
$\Delta x^- \sim 1/b\Lambda$ and $\Delta x^+ \sim \Lambda/Q^2$. Since,
by assumption, $b\ll 1$, we conclude that the semi--fast gluon cannot
discriminate the internal structure of the interaction region in Fig.
\ref{LCgeom}, as anticipated.

This feature drastically simplifies the calculation of the semi--fast
propagator. To better appreciate that, let us analyze first the
corresponding free propagator\footnote{The component $G_0^{--}$ is not
shown since it is irrelevant for the present purposes.} :
 \be\label{LCPROP} G_0^{ij}(p)=\delta^{ij}G_0(p),\quad
 G_0^{i-}(p)=\frac{p^i}{p^++i\epsilon}\,G_0(p),\quad
G_0^{-i}(p)=\frac{p^i}{p^+-i\epsilon}\,G_0(p)
 \,,\ee where
$G_0(p)=1/(2p^+p^--\bm{p}^2+i\epsilon)$ is the same as the propagator
of a free scalar field. Consider the scalar propagator, and perform the
Fourier transform to coordinate space:
 \be\label{G0X} G_0(x-y)&=&-i\int\limits_{\rm strip}
 \frac{d p^+}{2\pi}\,
 {\rm e}^{\frac{-i p^+(x^--y^-)}{2 p^+}}\,  \big[
    \Theta(p^+) \Theta (x^+ - y^+)
    -\Theta(-p^+) \Theta (y^+ - x^+)
    \big]\nn &{}&\qquad\qquad\times\
    \int \frac{d^2\bm{p}}{(2 \pi)^2}\, \, {\rm e}^{i\bm{p}\cdot
 (\x-\y) -i\frac{\bm{p}^2}{2p^+}(x^+-y^+)}\,,\,\, \ee
where the integration over $p^-$ has been computed via contour
techniques and the integration over  $p^+$ is restricted to the strip
$\Lambda \gg |p^+| \gg b\Lambda$. The interesting region is where both
$x^--y^-$ and $x^+-y^+$ are small compared to the typical variation
scales, $1/p^+$ and $1/p^-={2p^+}/{\bm{p}^2}$, respectively. Then we
can neglect the longitudinal and temporal coordinates within the
exponentials and perform the remaining integrations, to obtain:
 \be\label{G02} iG_0^{i j}(x-y)
  \simeq \delta^{i j} \,\,
    \frac{1}{4 \pi}
    \ln\frac{1}{b}\,\,
    \delta^{(2)}(\bm{x}-\bm{y}),
 \ee
where the logarithm has been generated by the restricted integration
over $p^+$ :
 \be\int\limits_{\rm strip}
 {\frac{dp^+}{2\pi}}\,   \frac{\Theta(p^+)}{2p^+}\, =\,
 - \int\limits_{\rm strip}
 \frac{dp^+}{2\pi}\,   \frac{\Theta(-p^+)}{2p^+}
    =\frac{1}{4 \pi}\,
    \ln\frac{1}{b}\,.\ee
The final result, Eq.~(\ref{G02}), is the same as the Fourier transform
of the simplified momentum--space propagator in which $\bm{p}^2$ is
neglected compared to $p^+p^-$ :
 \be\label{G03} G_0(x-y)
  \simeq \int\limits_{\rm strip}
 \frac{dp^+}{2\pi}\,\int \frac{dp^-}{2\pi}\,
 \frac{1}{2p^+p^-+ i\epsilon}
 \int \frac{d^2\bm{p}}{(2 \pi)^2}\, \,
 {\rm e}^{i\bm{p}\cdot (\x-\y)}\,.\ee
This is intuitive since small values for $x^+$ and $x^-$ correspond in
momentum space to large values for $p^-$ and $p^+$, respectively.
Similar approximations can be performed also in the presence of the
background fields.

Consider first the case where there is only one such a field, namely
$\delta\!A^-$. In the LC--gauge $a^+=0$, insertions of $\delta\!A^-$
cannot couple the transverse components $G^{ij}$ to any other component
like $G^{i-}$; thus, the calculation of $G^{ij}\,[\delta\!A^-]$ is
tantamount to that of the respective scalar propagator. The general
expression for $G^{ij}(x,y)[\delta\!A^-]$ valid for arbitrary points
$x$ and $y$ can be inferred from previous results in the literature
(see, e.g., Ref. \cite{RGE}). Here, however, we only need its
expression valid within the interaction region, and this can be easily
computed by solving the following differential equation
 \be\label{EikW} 2\del^+_x \big[\del^-_x - ig \delta\!A^-(x)\big]G(x,y) =
 \delta^{(4)}(x-y),\ee
which has been obtained from the general Dyson equation after
neglecting the transverse Laplacian $\grad_\perp^2$ relative to $\del^+
D^-$, so like in Eq.~(\ref{G03}). Working in the $p^+$ representation
(since $\delta\!A^-$ is independent of $x^-$) and using $D^-_x
W_{x^{+}y^{+}}(\bm{x})=0$ for $x^{+}>y^{+}$, where
\begin{align}\label{WilsonWF}
    W_{x^{+}y^{+}}(\bm{x})\equiv
    {\rm P} \exp\left\{{i g \int_{y^+}^{x^+}
    \dif z^+ \delta\!A^-(z^+,\bm{x})}\right\},
\end{align}
one immediately finds that the solution to (\ref{EikW}) with Feynman
boundary conditions reads
 \begin{align}\label{G1}
    iG^{i j}(x^+,\bm{x},y^+,\bm{y},p^+)[\delta\!A^-]
     =  \delta^{i j} \,
    \frac{1}{2 p^+}\,
    \delta^{(2)}(\bm{x}-\bm{y})
    \big[
    &\Theta(p^+) \Theta (x^+ - y^+) W_{x^+y^+}(\bm{x})
    \nonumber\\
    &-\Theta(-p^+) \Theta (y^+ - x^+) W_{y^+x^+}^{\dagger}(\bm{x})
    \big].
\end{align}
After also performing the restricted integration over $p^+$, one
finally obtains :
\begin{align}\label{G2}
    iG^{i j}(x^+,\bm{x},y^+,\bm{y})[\delta\!A^-] = \delta^{i j} \,
    \frac{1}{4 \pi}
    \ln\frac{1}{b}\,\,
    \delta^{(2)}(\bm{x}-\bm{y})
    \big[&\Theta (x^+ - y^+) W_{x^+y^+}(\bm{x})
    \nonumber \\
    &+\Theta (y^+ - x^+) W_{y^+x^+}^{\dagger}(\bm{x})
    \big].
\end{align}

Consider similarly the case where the only background field is the one
created by $\rho$, which is independent of $x^+$. If the field is taken
in the Coulomb gauge, where it has just a plus component $\tilde
B^+\equiv\alpha$, then the corresponding propagator is obtained by
simply replacing $x^+ \to x^-$ and  $\delta\!A^-\to \alpha$ (and
therefore  $W_{x^{+}y^{+}}(\bm{x}) \to U^{\dagger}_{x^-y^-}(\bm{x})$)
in Eq.~(\ref{G2}):
 \begin{align}\label{GU}
    i\wt{G}^{i j}(x^-,\bm{x},y^-,\bm{y})[\alpha] = \delta^{i j} \,
    \frac{1}{4 \pi}
    \ln\frac{1}{b}\,\,
    \delta^{(2)}(\bm{x}-\bm{y})
    \big[&\Theta (x^- - y^-) U^{\dagger}_{x^-y^-}(\bm{x})
    \nonumber \\
    &+\Theta (y^- - x^-) U_{y^-x^-}(\bm{x})
    \big].
\end{align}
Note that, {\em a priori}, the insertions of $\alpha$ can couple the
transverse semi--fast gluons $a^i$ to the temporal ones $a^-$, so in
general the propagator $\wt{G}^{ij}[\alpha]$ will involve a mixture of
all the components of the free propagator shown in Eq.~(\ref{LCPROP}).
Still, for small $x^-$ and $y^-$ (within the interaction region) the
coupling between transverse and temporal components is negligible since
the off--diagonal propagators like $G^{i-}_0$ are suppressed when $
p^+\gg p^i$.

If one now rotates the propagator (\ref{GU}) to the background
LC--gauge, $\wt{G}^{ij}[\alpha] \to {G}^{ij}[B^i]$ with
${G}^{ij}(x,y)=U(x)\wt{G}^{ij}(x,y)U^\dagger(y)$, cf. Eq.~(\ref{Grot}),
then one discovers that within the present approximations
${G}^{ij}[B^i]$ is the same as the free propagator (\ref{G02}) ! This
can be easily understood: Unlike the Coulomb gauge field $\alpha$ which
is singular at $x^-=0$ (on the resolution scale of the semi--fast
gluons), the LC--gauge field $B^i$ is only discontinuous there, so the
corresponding propagator ${G}^{ij}[B^i]$ must be continuous at both
$x^-=0$ and $y^-=0$. Thus, for small separations $x^{-}-y^{-} \simeq
0$,  the propagator will be independent of $x^-$ and $y^-$, and
therefore also independent of the field $B^i$.

Armed with this experience from the simpler cases, we now turn to the
general situation where both types of background field are
simultaneously present. We shall first take these fields in the
(background) LC--gauge, where $A^\mu=\delta^{\mu i}B^i+ \delta^{\mu
-}\delta\!A^-$. Then the propagator ${G}^{ij}[B^i,\delta\!A^-]$ will be
continuous at small $x^-$ (and thus independent of $B^i$), but
discontinuous at small $x^+$ (and thus dependent upon $\delta\!A^-$ via
the corresponding Wilson line (\ref{WilsonWF})). We conclude that the
general propagator that we need for the present purposes is the same as
the propagator (\ref{G2}) in the presence of $\delta\!A^-$ alone. This
is the propagator that we shall use for our most general calculations
below.

But it is still interesting (in particular, for the self--duality
argument in Sect. \ref{S_SEFF}) to also consider the general propagator
in the background Coulomb gauge, where  $\wt A^\mu=\delta^{\mu +}\alpha
+ \delta^{\mu -}\wt{\delta\!A}^-$. This is obtained from Eq.~(\ref{G2})
via the gauge--rotation (\ref{Grot}), and it is found to involve
products of temporal and longitudinal Wilson lines, of the type:
 \be\label{WUUW}
 \widetilde{W}_{x^{+}y^{+}}(x^-, \bm{x})
 U^{\dagger}_{x^-y^-}(y^{+},\bm{x}) \,=\,
  U^{\dagger}_{x^-y^-}(x^{+},\bm{x})
  \widetilde{W}_{x^{+}y^{+}}(y^-, \bm{x}),\ee
where $\widetilde{W}_{x^{+}y^{+}}$ is the temporal Wilson line in the
Coulomb gauge:
 \be\label{WCOV}
 \widetilde{W}_{x^{+}y^{+}}(x^-, \bm{x}) &\equiv&
 U^{\dagger}(x^-, x^{+},\bm{x})\,{W}_{x^{+}y^{+}}(\bm{x})\,
 U(x^-, y^{+},\bm{x})\,\nn &=&
    {\rm P} \exp\left\{ i g \int_{y^+}^{x^{+}}
    \dif z^+\, \widetilde{\delta\!A}^-(z^+,x^-,\bm{x})\right\},
 \ee
with $\widetilde{\delta\!A}^-$ the corresponding field, cf.
Eq.~(\ref{datilde}). The equality in Eq.~(\ref{WUUW}) follows because,
in the LC--gauge, ${W}_{x^{+}y^{+}}$ is independent of $x^-$, by
Eq.~(\ref{FPM}). Note that Eq.~(\ref{WUUW}) involves two possible paths
joining the points $(x^-,x^{+})$ and $(y^-,y^{+})$; both these paths
give the same contribution because of the condition $F^{+-}=0$, which
reflects the separation of scales in the problem. But by the same
condition, {\em any} other path joining these points will give an
identical contribution, so the expressions in Eq.~(\ref{WUUW}) can be
equivalently replaced by
 \be\label{gamma}
 {\rm P} \exp\left\{ i g \int_{\gamma} \big( \dif z^+\,
 \widetilde{\delta\!A}^-
 + \dif z^-\alpha \big)(z^+, z^-, \bm{x})\right\},
 \ee
where the path $\gamma$ going from $(y^-,y^{+})$ to $(x^-,x^{+})$ must
lie inside the interaction region, but otherwise is arbitrary. With the
Wilson line (\ref{gamma}), the Coulomb gauge propagator
$\wt{G}^{ij}[\alpha, \wt{\delta\!A}^-]$ is clearly invariant under the
transformations (\ref{DUAL1}), as anticipated in Sect. \ref{S_SEFF}.

\section{The high density regime: JIMWLK Hamiltonian}
\setcounter{equation}{0} \label{S_JIM}

Before we address the general case in the next section, it is
instructive to see how the formalism works in a case for which the
final result is already known: the strong field/no gluon number
fluctuation regime, where the RG evolution is governed by the JIMWLK
Hamiltonian \cite{JKLW97,RGE,W}. For more variety in the presentation,
the path that we shall follow in this section to recover the JIMWLK
Hamiltonian will be different from that to be used in the next section
in relation with the general case.

As it should be clear from a comparison between the diagrams in Figs.
\ref{JBfig}.a and c, and also from the general discussion in Sect.
\ref{S_RGE}, the effective action corresponding to this regime is
obtained by retaining terms to quadratic order in $\delta\!A^-$ and to
all orders in $B^i$ (or $\alpha$) in the general expressions
(\ref{Seff2}) and (\ref{Seff21}). Since the vertices $D^-F^{+ i}$ or
$D^+F^{- i}$ in these expressions start already at linear order in
$\delta\!A^-$, it will be enough for the present purposes to consider
the expression of the transverse propagator for $\delta\!A^-=0$. It
turns out that the JIMWLK Hamiltonian is most efficiently obtained
starting with Eq.~(\ref{Seff21}) and working in the background Coulomb
gauge. That is, we shall start with
\begin{align}\label{Seffdual}
    \Delta S_{\rm eff} = -\frac{1}{2} \int\limits_{x y}
    (2 \wt{D}^+ \wt{F}^{- i})_x
    \wt{G}^{i j}(x,y)
    (2 \wt{D}^+ \wt{F}^{- i})_y,
\end{align}
where $\wt D^+ = \del^+ -ig\alpha$, $\wt F^{- i}=-\del^i\wt{\dA}^-$,
and the propagator $\wt{G}^{i j}$ is given by Eq.~(\ref{GU}).

The first step is to integrate by parts over the longitudinal
coordinates $x^-$ and $y^-$. From Eq.~(\ref{GU}), one can easily check
that $\wt{D}^+_x \wt{G}^{i j}(x,y)=0$. By using this, together with the
identity ($A$ and $B$ are two generic color matrices)
 \begin{align}\label{id}
    \del^{\mu}(A B)= (D^{\mu}A)B + A (D^{\mu} B).
\end{align}
one can immediately deduce that the integrand in Eq.~(\ref{Seffdual})
is a total derivative w.r.t. both  $x^-$ and $y^-$, so the result of
the integration comes from the `surface' terms at large (positive or
negative) longitudinal coordinates. Specifically,
\begin{align}\label{Shigh0}
  \Delta  S_{\rm eff}
    = \frac{i}{2\pi}&\,\ln\frac{1}{b}
    \int\limits_{x^+,y^+,\bm{x}} \big[\wt F^{- i}_{x^+}
    (\infty,\bm{x})
    \wt F^{- i}_{y^+}(\infty,\bm{x})
    +\wt F^{- i}_{x^+}(-\infty,\bm{x})
    \wt F^{- i}_{y^+}(-\infty,\bm{x})
    \nonumber\\
    &-\wt F^{- i}_{x^+}(\infty,\bm{x})
    V^{\dagger}(\bm{x})
    \wt F^{- i}_{y^+}(-\infty,\bm{x})
    -\wt F^{- i}_{x^+}(-\infty,\bm{x})
    V(\bm{x})
    \wt F^{- i}_{y^+}(\infty,\bm{x}) \big],
\end{align}
where (cf. Eq.~(\ref{Utau1})) :
 \be\label{Vdef}
V^\dagger({\bm{x}})=\,{\rm P}\, {\rm exp}\left\{ig\int \dif x^-
\alpha^a(x^-,\bm{x})T^a\right\}\,.
 \ee
Note that the time arguments of the fields in Eq.~(\ref{Shigh0})
are shown as lower subscripts, a convention that we shall often
use in what follows. As explained in Sect. \ref{S_SEFF}, the
difference between the fields $\wt F^{- i}(x^-=\infty)$ and $\wt
F^{- i}(x^-=-\infty)$ comes from the interaction with the color
source $\rho$, as encoded in the gauge rotations in
Eq.~(\ref{datilde}). This interaction is localized at small $x^-$,
so $\wt F^{- i}(x^-)$ looks like a smooth $\Theta$--function in
$x^-$.

By using $\wt{F}^{-i}=-\del^i \wt{\dA}^-$ and defining
$\wt{A}^-(\vec{x})=\int \dif x^+\, \wt{\dA}^-(x)$, we obtain
\begin{align}\label{Shigh}
    \Delta S_{\rm eff}
    = \frac{i}{2\pi}&\,\ln\frac{1}{b}
    \int\limits_{\bm{x}} \big[\del^i \wt{A}^-(\infty,\bm{x})
    \del^i \wt{A}^-(\infty,\bm{x})
    +\del^i \wt{A}^-(-\infty,\bm{x})
    \del^i \wt{A}^-(-\infty,\bm{x})
    \nonumber\\
    &-\del^i \wt{A}^-(\infty,\bm{x})
    V^{\dagger}(\bm{x})
    \del^i \wt{A}^-(-\infty,\bm{x})
    -\del^i \wt{A}^-(-\infty,\bm{x})
    V(\bm{x})
    \del^i \wt{A}^-(\infty,\bm{x}) \big].
\end{align}
The evolution Hamiltonian can be now obtained via the replacement
(\ref{tilatorho}) (in four--dimensional coordinates !). By also using
the Poisson equation  (\ref{WTA}) to reexpress ${\delta}/{\delta
\wt\rho}$ in terms of ${\delta}/{\delta \alpha}$, we finally arrive at
\begin{align}\label{HJIMWLK}
    H_{\rm JIMWLK} =&
    \frac{-1}{(2\pi)^3}\,
    \int\limits_{\bm{x}\bm{y}\bm{z}}
    \mathcal{K}_{\bm{x}\bm{y}\bm{z}}\,
    \bigg[\frac{\delta}{\delta \alpha^a(\infty,\bm{x})}
    \frac{\delta}{\delta \alpha^a(\infty,\bm{y})}
    +\frac{\delta}{\delta \alpha^a(-\infty,\bm{x})}
    \frac{\delta}{\delta \alpha^a(-\infty,\bm{y})}
    \nonumber\\
    &-\frac{\delta}{\delta \alpha^a(\infty,\bm{x})}
    V_{ab}^{\dagger}(\bm{z})
    \frac{\delta}{\delta \alpha^b(-\infty,\bm{y})}
    -\frac{\delta}{\delta \alpha^a(-\infty,\bm{x})}
    V_{ab}(\bm{z})
    \frac{\delta}{\delta \alpha^b(\infty,\bm{y})} \bigg],
\end{align}
with the following transverse kernel:
 \be\label{Kdef} {\mathcal
K}({\bm{x}, \bm{y}, \bm{z}}) \,\equiv \,
   \frac{(\bm{x}-\bm{z})\cdot(\bm{y}-\bm{z})}{
     (\bm{x}-\bm{z})^2 (\bm{z}-\bm{y})^2}\,.\ee
In writing Eq.~(\ref{HJIMWLK}) we have  suppressed the $x^+$ dependence
although, strictly speaking, the integrand does depend upon time, via
the functional derivatives which should be interpreted as, e.g.,
 \be
\frac{\delta}{\delta \alpha^a(\infty,\bm{x})}\,\equiv\,\int \dif x^+\,
 \frac{\delta}{\delta \alpha^a(x^+,\infty,\bm{x})}\,.\ee
But this dependence is rather trivial since when the Hamiltonian acts
on scattering observables, like Eq.~(\ref{up}), the only contribution
arises from the interaction time $x^+\simeq 0$.

On the other hand,  the $x^-$ dependence of the Hamiltonian
(\ref{HJIMWLK}) is less trivial and deserves some comments: As
mentioned before, $\wt{\dA}^-(x)$ looks like a smooth
$\Theta$--function in $x^-$, which varies within the range  $0
\simle x^- \simle x^-_\tau$, with $x^-_\tau\sim 1/\Lambda =
(1/P^+) \rme^\tau$. Therefore, in the identification
(\ref{tilatorho}), the asymptotic fields
$\wt{\dA}^-(x^-\to\pm\infty)$ should be really understood as the
values of the field at the end points $x^-_0 \simeq 0$ and
$x^-_\tau$ of this finite interval. Thus, strictly speaking,
  \be\label{alphaINFTAU}
\frac{\delta}{\delta \alpha^a(\infty,\bm{x})}\,\equiv\,
\frac{\delta}{\delta \alpha^a_\tau(\bm{x})}\qquad{\rm and}\qquad
\frac{\delta}{\delta \alpha^a(-\infty,\bm{x})}\,\equiv\,
 \frac{\delta}{\delta \alpha^a_0(\bm{x})},\ee
act as derivatives at the end points of the Wilson line in
Eq.~(\ref{Utau1}):
 \be\label{Lie}
 \hspace*{-3mm}\frac{\delta
V^\dag({\x})}{\delta \alpha_{\tau}^a(\y)}=ig\delta^{(2)}(\x-\y)
T^aV^\dag({\x})\,, \qquad \frac{\delta V^\dag({\x})}{\delta
 \alpha_{0}^a(\y)}=ig\delta^{(2)}(\x-\y)V^\dag({\x})T^a.\ee
By comparing these results and using the identity $V_{ab}^{\dagger}T^b=
V T^a V^{\dagger}$, one can express the derivatives at $x^-_0$ in terms
of those at $x^-_\tau$:
\begin{align}\label{apminf}
    \frac{\delta}{\delta \alpha^{a}_0(\bm{x})}\,=\,
    \frac{\delta}{\delta \alpha^{b}_{\tau}(\bm{x})}\,
    V_{ba}^{\dagger}(\bm{x}) \,=\,
    V_{ab}(\bm{x})\,
    \frac{\delta}{\delta \alpha^{b}_{\tau}(\bm{x})}\,.
 \end{align}
Alternatively, and more generally, Eq.~(\ref{apminf}) can be
obtained as the result of a gauge rotation, as we show now. Let us
first introduce the gauge--covariant `color current' :
 \be\label{defJ-}
  J^-(x)\equiv D_{\nu}F^{\nu -}(x)= D_{i}F^{i-} + D_{+}F^{+-}
  = D^{i}F^{-i},\ee
where in writing the last equality we have used $F^{+-}=0$, cf.
Eq.~(\ref{FPM}). (In a scattering problem, in which $\delta A^-$
would be the field created by a left--moving projectile, $J^-_a$
would be the color current associated with the latter.) Under the
gauge rotation (\ref{WilsonU}) from the  LC-- to the Coulomb
gauge, this current transforms like:
  \be\label{WTJ-}
  \wt J^-(x)\,
 \,=\,U^{\dagger}(x)\,J^-(x)\,U(x)
 \,,\ee
a relation which for $x^-=\infty$ can be rewritten as:
 \be\label{WTJ1}
  \wt J^-_a(x^+,\infty,\bm{x})\,=\,U^{\dagger}_{ab}(x^+,\infty,\bm{x})
  \,\wt J^-_a(x^+,-\infty,\bm{x})
  \,.\ee
By also using $\wt{F}^{-i}=-\del^i \wt{\dA}^-$ together with
Eqs.~(\ref{defJ-}), (\ref{tilatorho}) and (\ref{WTA}), one can
successively write
 \be
 \wt J^-_a(x)\,=\,
 - \nabla^2_\perp \wt{\dA}^-_a(x)\ \longrightarrow\ - \nabla^2_\perp
  \,\frac{{i}\,\delta}{\delta \wt{\rho}_a(x)} \,=\, i\,
 \frac{\delta}{\delta \alpha_a(x)}\,.\ee
By inserting this representation for $\wt J^-_a(x)$ in both the
l.h.s. and the r.h.s. of Eq.~(\ref{WTJ1}), and recalling that
$U^{\dagger}(x^+,\infty,\bm{x})\approx V^{\dagger}(\bm{x})$ in the
present regime (small ${\dA}^-$), we finally recover the relation
(\ref{apminf}), as anticipated.

After using Eqs.~(\ref{alphaINFTAU}) and Eq.~(\ref{apminf}), the JIMWLK
Hamiltonian (\ref{HJIMWLK}) takes its standard form:
 \begin{align}\label{HJIM}
    H_{\rm JIMWLK} =&
    \frac{-1}{(2\pi)^3}\,
    \int\limits_{\bm{x}\bm{y}\bm{z}}
    \mathcal{K}_{\bm{x}\bm{y}\bm{z}}\,
    \frac{\delta}{\delta \alpha^{a}_{\tau}(\bm{x})}
    \Big[1 +V^\dag_{\x}
    V_{\y}- V^{\dagger}_{\z} V_{\y} -V^{\dagger}_{\x} V_{\z}
    \Big]^{ab}
    \frac{\delta}{\delta \alpha^{b}_{\tau}(\bm{y})}
    ,
\end{align}
as originally derived in Ref. \cite{RGE}. Note that, although we
have explicitly computed here only the `real' quantum correction,
i.e., the gluon exchange diagram in Fig. \ref{JBfig}.a, the above
Hamiltonian is complete as written, that is, it also contains the
`virtual' correction (associated with self--energy and vertex
corrections), which is generated when commuting the functional
derivatives in Eq.~(\ref{HJIM}) through the Wilson lines in the
kernel there \cite{W,RGE}. The proper ordering of the operators
inside the Hamiltonian (which implicitly takes into account the
virtual corrections) has naturally emerged from the previous
manipulations because of the intimate connection between the
ordering of the operators and the gauge symmetry : The operators
in Eqs.~(\ref{HJIMWLK}) or (\ref{HJIM}) are line--ordered in $x^-$
in such a way that the Hamiltonian be invariant under the gauge
transformations which depend only upon $x^-$ (the
residual\footnote{These are the transformations which preserve the
structure of the classical field  in the Coulomb gauge: $\tilde
B^\mu=\delta^{\mu +}\alpha$ with $\alpha$ independent of $x^+$ ;
see Ref. \cite{ODDERON} for details.} gauge transformations in the
Coulomb gauge).

Let us conclude this section with some considerations about the
Hamiltonian structure associated with the JIMWLK Hamiltonian (see
also Refs. \cite{RGE,W,PATH,ODDERON}). Such considerations may
look unnecessarily formal at this level, but they will be useful
for clarifying more general cases later on. So far, we have
implicitly assumed that $H_{\rm JIMWLK}$ acts as a (second--order)
functional differential operator in the Hilbert space spanned by
the observables ${\cal O}(V,V^\dagger)$ built with the Wilson
lines. This is sufficient for the present purposes since, as we
have seen in Sect. 2, all the observables pertinent to the
scattering between the high--density right--moving target and a
(relatively simple) left--moving projectile are built in terms of
Wilson lines. Within this Hilbert space, the two functional
derivatives introduced in Eq.~(\ref{alphaINFTAU}) act,
respectively, as left and right Lie derivatives\footnote{Note that
the ``left'' and ``right'' nomenclature refers, by convention, to
the action of the Lie derivatives on $V^{\dagger}$ ; for the
corresponding action on $V$, the ``left'' and ``right'' would be
interchanged (see, e.g., Eq.~(\ref{LieO}) below).}
 (i.e., as generators of the infinitesimal
left and right gauge rotations). Specifically, if one writes
 \be\label{JLRdef}
J_{L}^a(\x) \equiv \frac{1}{i} \frac{\delta}{\delta
\alpha_\tau^a({\x})}\,,\qquad J_{R}^a(\x) \equiv \frac{1}{i}
 \frac{\delta}{\delta
 \alpha_0^a({\x})} \,=\,
    V_{ab}(\bm{x})\,J_{L}^b(\x)
    \,,\ee
then one has, for instance (with the Wilson lines taken in some
arbitrary representation of the color group, and the transverse
coordinates omitted, for simplicity),
 \be\label{LieO}\hspace*{-0.3cm}
 J_{L}^a \, V^{\dagger}_{ij}
 &\,=\,& g(T^a V^{\dagger})_{ij}\,,\qquad
 J_{L}^a \, V_{ij}
 \,=\, -g(VT^a )_{ij} \,,\nn
 J_{L}^a \,{\cal O}(V,V^\dagger)
 &\,=\,& g(T^a V^{\dagger})_{ij}\,\frac{\del{\cal O}}
 {\del V^{\dagger}_{ij}} \,-\,g(VT^a )_{ij}\,
 \frac{\del{\cal O}} {\del V_{ij}}\,,\ee
together with similar relations for the action of $J_{R}^a$. These
equations, as well as Eq.~(\ref{Lie}), can be rewritten as
commutators between formal operators:
 \be\label{LieFin}\hspace*{-0.3cm}
 \big[J_{L}^a(\x), \, V^{\dagger}_{bc}(\bm{y})\big]
 &\,=\,& g(T^a V^{\dagger}(\bm{x}))_{bc}\,\delta^{(2)}(\x-\y)\,,\nn
 \big[J_{R}^a(\x), \,V^{\dagger}_{bc}
 (\bm{y})\big]&\,=\,& g(V^{\dagger}(\bm{x})T^a)_{bc}\,\delta^{(2)}(\x-\y)\,,
 \ee
The left and right Lie derivatives span two independent SU($N_c$)
algebras :
 \be\label{commLR}
 \big [J_{L}^a(\x),\, J_{L}^b(\y) \big]
  &\,=\,& - if^{abc}J_{L}^c(\x) \,\delta^{(2)}(\x-\y)\,,\nn
 \big [J_{R}^a(\x),\, J_{R}^b(\y) \big] &\,=\,&  if^{abc}J_{R}^c(\x)
 \,\delta^{(2)}(\x-\y)\,, \nn
 \big[J_{L}^a(\x), \, J_{R}^b(\y)\big] &\,=\,& 0 .\ee
The first two equations above follow by enforcing the Jacobi
identity in Eq.~(\ref{LieFin}), whereas the last commutator,
namely
 \be\label{LR0}
\big[J_{L}^a(\x), \, J_{R}^b(\y)\big] = \big[J_{L}^a(\x), \,
V_{bc}(\bm{y})\big] J_{L}^c(\y) + V_{bc}(\bm{y})\big
 [J_{L}^a(\x),\, J_{L}^c(\y) \big]\,=\,0,\ee
vanishes because the infinitesimal gauge rotation  of the Wilson
line is compensated by the commutator of $J_{L}$ with itself.

We are now prepared to characterize the canonical Hamiltonian
structure associated with the JIMWLK evolution: The Wilson line
$V^{\dagger}_{bc}(\x)$ and the left Lie derivative $J_{L}^a(\x)$
play the roles of the canonical coordinate and its conjugate
momentum, respectively, and the Hamiltonian $H_{\rm
JIMWLK}[J_{L},V^{\dagger}]$ --- as defined by Eqs.~(\ref{HJIM})
and (\ref{JLRdef}) --- acts on the respective phase--space via the
Poisson brackets defined by the `left' part of the commutation
relations\footnote{Of course, the Poisson brackets of
$V^{\dagger}$ with itself, or with $V$, are trivial:
$[V^{\dagger}_{ab}(\x), V^{\dagger}_{cd}(\y)]=0$, etc.} in
Eqs.~(\ref{LieFin})--(\ref{commLR}). The evolution of an arbitrary
observable ${\cal O}(V,V^\dagger)$ is then governed by the
canonical equation of motion :
      \be
     \frac{\partial}{\partial \tau} \, {\cal O}(V,V^\dagger)= -
     \big[H_{\rm JIMWLK},\,{\cal O}\,
     \big] \, ,
  \label{EVOLCAN}
  \ee
which after averaging with the target weight function (now
expressed as a functional of $V$ and $V^\dagger$)  is equivalent
to Eq.~(\ref{RGEgen2}).

In particular, when acting on dipole scattering operators  so like
Eq.~(\ref{SdipoleA}), Eq.~(\ref{EVOLCAN}) produces the same
evolution equations as originally obtained by Balitsky \cite{B}
from the analysis of projectile evolution in a strong background
field. This is natural since the gluon mergings in the target
wavefunction, cf. Fig. \ref{JBfig}.a, can be reinterpreted as
splittings in the projectile wavefunction followed by multiple
scattering between the products of this splitting and the strong
target field. In fact, the effective action (\ref{Shigh}) can be
understood as describing one--step quantum evolution in the
scattering between a high--density target (represented by the
strong classical color field $\alpha$) and a low--density
projectile (the source of the weak field $\wt{A}^-$). With this
interpretation, Eq.~(\ref{Shigh}) is equivalent with the `real'
part of the effective action obtained by Balitsky  (see Eq.~(41)
in the second paper of Ref. \cite{B1}).

\section{The general effective action}
\setcounter{equation}{0} \label{S_GEN}

Turning to the general case in which both background fields are strong
(at least, formally), it turns out that the calculation is most
conveniently performed by working in the background LC--gauge and using
the expression (\ref{Seff2}) for the effective action. Indeed, the
corresponding background field propagator is explicitly known (cf.
Eq.~(\ref{G2})), and involves only one type of Wilson lines. Also, the
vertex $D^-F^{+ i}=D^-\del^+ B^i$ which appears in Eq.~(\ref{Seff2}) is
a total derivative (in a gauge--covariant sense) with respect to both
$x^+$ and $x^-$, so the corresponding integrations can be easily done.
In fact, the propagator (\ref{G2}) satisfies $D^-_x G^{ij}(x,y)=0$ and
$\del^+_x G^{ij}(x,y)=0$, so both integrations receive contributions
from the boundary terms alone.

After performing the integrations over $x^+$ and $y^+$, one finds:
\begin{align}\label{SF}
    \Delta S_{\rm eff}
    = &\frac{i}{2\pi}\,\ln\frac{1}{b}
    \int\limits_{x^-,y^-,\bm{x}} \big[F_{\infty}^{+i}(x^-,\bm{x})
    F_{\infty}^{+i}(y^-,\bm{x})
    +F_{-\infty}^{+i}(x^-,\bm{x})
    F_{-\infty}^{+i}(y^-,\bm{x})
    \nonumber\\
    &-F_{\infty}^{+i}(x^-,\bm{x})
    W_{\infty,-\infty}(\bm{x})
    F_{-\infty}^{+i}(y^-,\bm{x})
    -F_{-\infty}^{+i}(x^-,\bm{x})
    W^{\dagger}_{\infty,-\infty}(\bm{x})
    F_{\infty}^{+i}(y^-,\bm{x}) \big].
\end{align}
By using $F^{+i}=\del^{+} B^i$ and recalling that the field $B^i$
vanishes at $x^- = -\infty$, one can also perform the integrations over
$x^-$ and $y^-$, with the following result:
\begin{align}\label{SA}
   \Delta S_{\rm eff}
    = \frac{i}{2\pi}\,\ln\frac{1}{b}
    \int\limits_{\bm{x}} \,&\big[\mathcal{B}^i_{\infty}(\bm{x})
    \mathcal{B}^i_{\infty}(\bm{x})
    +\mathcal{B}^i_{-\infty}(\bm{x})
    \mathcal{B}^i_{-\infty}(\bm{x})
    \nonumber\\
    &-\mathcal{B}^i_{\infty}(\bm{x})
    W_{\infty,-\infty}(\bm{x})
    \mathcal{B}^i_{-\infty}(\bm{x})
    -\mathcal{B}^i_{-\infty}(\bm{x})
    W^{\dagger}_{\infty,-\infty}(\bm{x})
    \mathcal{B}^i_{\infty}(\bm{x}) \big].
\end{align}
where $\mathcal{B}_{x^+}^i(\bm{x})\equiv B^i(x^+,x^-=\infty,\bm{x})$.

As anticipated, the effective action (\ref{SA}) is completely
determined by the asymptotic values of the fields at large $x^+$ and
$x^-$, which in turn are the same as the respective fields along the
edges of the interaction region in Fig. \ref{LCgeom} (since the fields
remain constant outside this region). This property is essential for
the success of the renormalization group analysis, as it insures that
the separation of scales is preserved by the quantum evolution: The
effective dynamics at the soft scale `sees' only a coarse--grained (in
$x^+$ and $x^-$) version of the dynamics at faster scales, but not also
the short--range details of the latter.

Our final step will be to express the action fully in terms of Wilson
lines. To that aim, one should recall that the field $\mathcal{B}^i$ is
a pure gauge given by
\begin{align}\label{Apg}
    \mathcal{B}^i_{x^+}(\bm{x})= \frac{i}{g}\,
    V_{x^+} \del^i\, V^{\dagger}_{x^+},
\end{align}
where $V^{\dagger}=U^{\dagger}(x^-=\infty)$ with the Wilson line
$U^{\dagger}$ given in Eq.~(\ref{WilsonU}). The longitudinal
Wilson lines are explicitly known in terms of the field $\alpha$
in the Coulomb gauge, so it is convenient to fully rewrite
Eq.~(\ref{SA}) in this gauge. From Eq.~(\ref{WCOV}), we deduce
that
 \be\label{WilsonWt}
    \widetilde{W}_{\infty,-\infty}(\infty, \bm{x})&\, =\, &
    V^{\dagger}_{\infty}(\bm{x})
    W_{\infty,-\infty}(\bm{x})
    V_{-\infty}(\bm{x})\nn &\, =\, &
    {\rm P} \exp\left\{i g \int_{-\infty}^{\infty}
    \dif z^+\, \widetilde{\delta\!A}^-(z^+,x^-=\infty,\bm{x})\right\},
 \ee
whereas $W_{\infty,-\infty}(\bm{x})$ can be identified with the Coulomb
gauge temporal line at $x^-=-\infty$: $W_{\infty,-\infty}(\bm{x}) =
\widetilde{W}_{\infty,-\infty}(-\infty, \bm{x})$. Let us introduce the
simpler notations:
 \be\label{Wfinal}
 \widetilde{W}_{\infty,-\infty}(\infty,
 \bm{x})\,\equiv\,{W}_{\infty}(\bm{x}),\qquad
 \widetilde{W}_{\infty,-\infty}(-\infty, \bm{x})
 \,\equiv\,{W}_{-\infty}(\bm{x}).\ee
By using Eqs.~(\ref{Apg})--(\ref{Wfinal}) and some simple
manipulations, one can rewrite Eq.~(\ref{SA}) in terms of the Wilson
lines which delimitate the interaction region in Fig. \ref{LCgeom} :
\begin{align}\label{SVW}
    \Delta S_{\rm eff}
    = \frac{i}{2\pi g^2 N_c}\,\ln\frac{1}{b}
    \int\limits_{\bm{x}}
    \Tr
    &\big[(\del^i V_{\infty})
    (\del^i V^{\dagger}_{\infty})
    +(\del^i V_{-\infty})
    (\del^i V^{\dagger}_{-\infty})
    \nonumber\\
    &-(\del^i V^{\dagger}_{\infty})
    {W}_{-\infty}
        (\del^i V_{-\infty})
    {W}^{\dagger}_{\infty}
    -(\del^i V_{\infty})
    {W}_{\infty}
        (\del^i V^{\dagger}_{-\infty})
    {W}^{\dagger}_{-\infty}
    \big],
\end{align}
where we have also made use of the identity
\begin{align}\label{id2}
    W^{ab}= \frac{1}{N_c} \Tr[T^a W T^b W^{\dagger}].
\end{align}
Note that the lower subscripts $\pm\infty$ refer to $x^+$ in the
case of $V$ and $V^{\dagger}$, but to $x^-$ for $W$ and
$W^{\dagger}$ (see also Fig. \ref{LCgeom}). Let us now discuss
some general properties of the effective action (\ref{SVW}), with
emphasis on gauge symmetry and self--duality.

The action (\ref{SVW}) is invariant under $\bm{x}$--independent
gauge transformations, which are the residual transformations
permitted by the Coulomb gauge. Indeed, if $h_{x^+}(x^-)\in$
SU$(N_c)$ is such a transformation, then
 ${W}_{\pm\infty} \to h_{\infty}(\pm\infty) {W}_{\pm\infty}
h^{\dagger}_{-\infty}(\pm\infty)$, $V_{\infty} \to h_{\infty}(-\infty)
V_{\infty} h^{\dagger}_{\infty}(\infty)$ and $V^{\dagger}_{\infty} \to
h_{\infty}(\infty) V^{\dagger}_{\infty} h^{\dagger}_{\infty}(-\infty)$,
so that the trace in (\ref{SVW}) remains unchanged.

From the discussion in Sect. \ref{S_SEFF} we expect the expression
in Eq.~(\ref{SVW}) to be `self--dual', i.e., invariant under the
transformations (\ref{DUAL1}). However, this symmetry is not
manifest at the level of Eq.~(\ref{SVW}), where the temporal and
longitudinal Wilson lines enter on a different footing. This
asymmetry reflects our prior use of the background LC--gauge, in
which the gluon propagator involves only the temporal Wilson line.
However, after some integrations by parts, it is possible to
rewrite  Eq.~(\ref{SVW}) in such a way that self--duality becomes
manifest. We shall write, e.g., \be
 \int\limits_{\bm{x}}
    \Tr
    (\del^i V_{\infty})
    (\del^i V^{\dagger}_{\infty})&=&
 \frac{1}{2} \int\limits_{\bm{x}} \Tr[ -(\partial^2
 V_\infty) V^{\dagger}_{\infty}- V_\infty
 (\partial^2V^{\dagger}_{\infty})],\nn
 \int\limits_{\bm{x}}
    \Tr (\del^i V^{\dagger}_{\infty})
    {W}_{-\infty}
        (\del^i V_{-\infty})
    {W}^{\dagger}_{\infty}&=&
 \frac{1}{2} \int\limits_{\bm{x}} \Tr[
 V^{\dagger}_\infty
 \partial^i(W_{-\infty}(\partial^i V_{-\infty})W^{\dagger}_\infty)
  +\partial^i(W^{\dagger}_\infty(\partial^i
 V^{\dagger}_\infty)
 W_{-\infty}) V_{-\infty}],\nonumber\ee
and then repeatedly use Eqs.~(\ref{WilsonWt})--(\ref{Wfinal}) to
finally deduce two different but equivalent\footnote{The integrands in
the two following expressions are Hermitian conjugate to each other and
separately real, so they are indeed identical.} expressions for the
effective action:
\begin{align}\label{selfdual}
    \Delta S_{\rm eff}
    = \frac{i}{2\pi g^2 N_c}\,\ln\frac{1}{b}
    \int\limits_{\bm{x}}
    \Tr
    &\big[V^{\dagger}_\infty
    (\partial^iW_{-\infty})(\partial^i V_{-\infty})
    W^{\dagger}_\infty+ V^{\dagger}_\infty
    W_{-\infty}(\partial^i V_{-\infty})(\partial^i W^{\dagger}_\infty)
    \nonumber \\
    &+(\partial^iW^{\dagger}_\infty)(\partial^i
    V^{\dagger}_\infty)
    W_{-\infty} V_{-\infty}+W^{\dagger}_\infty(\partial^i
    V^{\dagger}_\infty)
   (\partial^i W_{-\infty}) V_{-\infty}\big],\end{align}
and respectively
\begin{align}\label{selfdual1}
    \Delta S_{\rm eff}
    = \frac{i}{2\pi g^2 N_c}\,\ln\frac{1}{b}
    \int\limits_{\bm{x}}
    \Tr
    &\big[
   V_\infty(\partial^i W_{\infty})(\partial^i
   V_{-\infty}^{\dagger})W^{\dagger}_{-\infty}+V_\infty
   W_{\infty}(\partial^i V_{-\infty}^{\dagger})
   (\partial^iW^{\dagger}_{-\infty}) \nonumber \\
   &+(\partial^iW^{\dagger}_{-\infty})(\partial^i V_\infty)
   W_{\infty} V_{-\infty}^{\dagger}+W^{\dagger}_{-\infty}(\partial^i V_\infty)
  (\partial^iW_{\infty}) V_{-\infty}^{\dagger}
    \big].
\end{align}
It is likely that this `degeneracy' in the form of the effective action
corresponds to the two original versions of it, Eqs.~(\ref{Seff2}) and
(\ref{Seff21}), respectively. In any case, it is now easy to check by
inspection that Eqs.~(\ref{selfdual}) and (\ref{selfdual1}) get
interchanged with each other under the following transformations
\begin{align}\label{DUALWV} W_{\infty}
 \longleftrightarrow V_{\infty}^{\dagger},  \qquad W_{-\infty}
 \longleftrightarrow V_{-\infty}^{\dagger}, \end{align}
which represent a stronger version of the transformations (\ref{DUAL1})
in which the ordering of the Wilson lines is also specified. Thus, the
effective action is invariant under the duality transformation (or
`self--dual'), as anticipated.

Eq.~(\ref{selfdual}) has other interesting symmetries. For example, it
is invariant under the following transformation
  \begin{align} W_{-\infty} \to V_{-\infty},  \qquad
   V_{-\infty} \to W_{\infty}^{\dagger},  \qquad
   W_{\infty}^{\dagger} \to V_{\infty}^{\dagger},  \qquad
   V_{\infty}^{\dagger} \to W_{-\infty},
    \end{align}
which can be recognized as the clockwise $90^{\rm o}$ rotation of the
diamond--shaped interaction area in Fig. \ref{LCgeom}.

Note also that Eqs.~(\ref{WilsonWt})--(\ref{Wfinal}) imply $W_{\infty}=
V^{\dagger}_{\infty} W_{-\infty} V_{-\infty}$, which can be rewritten
as:
 \be\label{diamond}
V^{\dagger}_{\infty} W_{-\infty}
V_{-\infty}W_{\infty}^{\dagger}\,=\,1 \ \Longleftrightarrow \
\frac{1}{N_c^2-1}\,{\rm Tr}\, W_{\Diamond}\,=\,1,\ee where
$W_{\Diamond}$ is the Wilson loop built around the interaction
`diamond' in Fig. \ref{LCgeom}. The fact that the overall Wilson
loop is trivial is a consequence of the condition $F^{+-}=0$ which
expresses the separation of scales in the problem. (In fact,
Eq.~(\ref{diamond}) is still another gauge--invariant expression
of this separation of scales.) Eq.~(\ref{diamond}) can be also
recognized as the generalization of Eq.~(\ref{apminf}) to the case
where both background fields are strong. Indeed, as we shall
shortly discuss,  $W_{\infty}$ and $W_{-\infty}$ play the same
role in the general Hamiltonian as the functional Lie derivatives
(\ref{alphaINFTAU}) in the JIMWLK Hamiltonian. Because of the
condition $F^{+-}=0$, the (generalized) derivatives at the `end
points' $x^-=\infty$ and $x^-=-\infty$ are not independent of each
other, but rather they are related by a gauge rotation, as
manifest in either Eq.~(\ref{apminf}), or Eq.~(\ref{diamond}).

By exploiting the self--duality property, one can deduce one more
form for the effective action, which will be useful for comparison
with the JIMWLK action in the dilute regime. Specifically, by
applying the duality transformations (\ref{DUALWV}) to
Eq.~(\ref{SVW}), one finds
\begin{align}\label{SVIW}
    \hspace*{-0.3cm}
    \Delta S_{\rm eff}
    = \frac{i}{2\pi g^2 N_c}\,\ln\frac{1}{b}
    \int\limits_{\bm{x}}
    \Tr
    &\big[(\del^i W^{\dagger}_{\infty})
    (\del^i W_{\infty})
    +(\del^i W^{\dagger}_{-\infty})
    (\del^i W_{-\infty})
    \nonumber\\
    &-(\del^i W_{\infty})
    V^{\dagger}_{-\infty}
        (\del^i W^{\dagger}_{-\infty})
    V_{\infty}
    -(\del^i W^{\dagger}_{\infty})
    V^{\dagger}_{\infty}
        (\del^i W_{-\infty})
    V_{-\infty}
    \big].
\end{align}

Any of the equivalent equations (\ref{SVW}), (\ref{selfdual}),
(\ref{selfdual1}), and  (\ref{SVIW}) defines a two--dimensional
field theory in the space spanned by the asymptotic Wilson lines,
with a local action. Because of the condition (\ref{diamond}),
only three among the four Wilson lines which enter any of these
expressions should be treated as independent field variables;
these can be chosen, e.g., as $V^{\dagger}_{\infty}$,
$V_{-\infty}$, and $W_{\infty}^{\dagger}$.

In order to complete the construction of the Hamiltonian theory,
one still needs to specify the commutation relations among the
independent Wilson lines. In principle, these could be deduced by
using the expressions of the Wilson lines as path--ordered
exponentials of $\alpha_a(x)$ and ${\delta}/\delta
\wt{\rho}_a(x)$, respectively, together with the elementary
commutation relations among these Lie--valued fields and
derivatives; e.g., the Poisson equation (\ref{WTA}) implies:
 \be\label{commut0}
 \left[\frac{\delta}{\delta \wt{\rho}_a(x)},\, \alpha_b(y) \right]
 \,=\,\delta_{ab}\,\delta(x^+-y^+)\,\delta(x^--y^-)\, \Delta(\x-\y)\,,\ee
with $\Delta(\x-\y)$ the two--dimensional Coulomb propagator
introduced in Eq.~(\ref{alpha}). In fact, in the relevant Wilson
lines, the field $\alpha_a(x)$ appears only for asymptotic values
of time, $x^+= \pm\infty$; similarly, ${\delta}/{\delta
\wt{\rho}_a(x)}$ enters only for $x^-= \pm\infty$ (which truly
means $x^-_\tau$ and $x^-_0$ in the notations of
Eq.~(\ref{alphaINFTAU})). Accordingly, the commutation relations
among $V^{\dagger}_{\infty}$ and $W_{\infty}^{\dagger}$ should
receive non--trivial contributions only from their {\em end
points} at $x^+=x^-=\infty$.

But the construction of the general commutation relations is
complicated by the fact that, e.g., the end--point derivatives
${\delta}/{\delta \alpha_\tau^a}$ do not commute with each other,
but rather obey the SU($N_c$) algebra in Eq.~(\ref{commLR}); via
Eq.~(\ref{WTA}), this implies non--trivial commutators among the
derivatives ${\delta}/{\delta \wt{\rho}^{\,a}_\infty}$. In the
next section we shall discover that, in the dilute regime, the
field variables $\alpha_\infty^a$ should be treated as
non--commuting variables (and similarly for $\alpha_{-\infty}^a$)
\cite{KL05}. Hence, in the general case, the (end--point) gauge
fields do not commute with themselves, neither do so the
respective functional derivatives. If taken at face value, such
non--commutativity properties at the level of the Lie--valued
fields and derivatives would imply that a Wilson--line commutator
like
$$[V^{ab}_{\infty}(\bm{x}),
V^{cd}_{\infty}(\bm{y})]$$ is not only  non--vanishing, but also
non--local in $x^-$ ! This would spoil the two--dimensional
character of the effective theory at high energy. We do not know
yet what could be the solution to this difficulty in the general
case (if any !), yet the fact that its two limiting cases --- the
high--density regime, cf. Sec.~\ref{S_JIM}, and the low--density
one, to be discussed in Sec.~\ref{S_BREM} --- can be unambiguously
formulated as two--dimensional, Hamiltonian, field theories
strongly suggests that a similar construction should exist also in
the general case. We postpone this issue for latter
investigations.

Let us finally check that, as anticipated at several places, the
general effective action reduces to the JIMWLK action in the
high--density regime where the fluctuations become unimportant and
the radiated field $\wt{\dA}^-$ can be treated as weak. In this
regime, the temporal Wilson lines $W_{\infty}$ and $W_{-\infty}$
can be expanded out in  perturbation theory, whereas the
time--dependence of the longitudinal Wilson lines can be ignored:
$V^{\dagger}_{-\infty}\approx V^{\dagger}_{\infty} \equiv
V^{\dagger}$, etc. To study this limit, it is convenient to start
with the form (\ref{SVIW}) for the effective action, since there
all the temporal Wilson lines appear under transverse gradients.
Thus, to obtain the expansion of Eq.~(\ref{SVIW}) to quadratic
order in $\wt{\dA}^-$, it suffices to expand each temporal line to
linear order:
 \be
  \del^i W^{\dagger}_{\infty}\approx -ig\,
  \del^i\wt{A}^-(\infty,\bm{x}),\qquad \del^i W_{-\infty}
  \approx ig\, \del^i\wt{A}^-(-\infty,\bm{x}), \ee
where $\wt{A}^-(x^-,\x)\equiv \int \dif x^+\, \wt{\dA}^-(x)$.
Then, clearly, Eq.~(\ref{SVIW}) reduces to Eq.~(\ref{Shigh}),
which is the effective action corresponding to the JIMWLK
Hamiltonian. In the next section, we shall similarly consider the
limit of the general action in the dilute regime, and construct
the associated Hamiltonian theory.

\section{The low density regime: Bremsstrahlung Hamiltonian}
\setcounter{equation}{0} \label{S_BREM}

In the case of a dilute target, the color field $\alpha$ is weak
and the longitudinal Wilson lines in the effective action can be
expanded in perturbation theory. If one starts with
Eq.~(\ref{SVW}), it is sufficient to consider this expansion to
lowest order, e.g.,
 \be
  \del^i V^{\dagger}_{\pm\infty}(\x)\approx ig\,\int \dif
  x^-\,\del^i\alpha_{\pm\infty}^a(x^-,\x)T^a\,=\,\frac{-ig}{2\pi}
    \int\limits_{\bm{z}}\frac{(\bm{x}-\bm{z})^i}
    {(\bm{x}-\bm{z})^2}\,\wt\rho_{\pm \infty}^{\,a}(\bm{z})T^a,\ee
where we have also used the equation (\ref{WTA}) relating the
field $\alpha$ to the Coulomb--gauge color source $\wt\rho$. In
the last equality above, $\wt\rho(\bm{z})$ stands for the color
charge density integrated over the longitudinal coordinate $x^-$.
Within the same approximations, one can neglect the difference
between quantities in the Coulomb and the LC gauges, and also the
$x^-$--dependence of the temporal Wilson lines (since the gauge
rotations in equations like (\ref{WilsonWt}) become negligible for
weak $\alpha$); that is, $W_{-\infty}\approx W_{\infty} \equiv W$.
It is then straightforward to derive the following limiting form
for the effective action \cite{KL05,KL3} :
\begin{align}\label{Slow}
   \Delta S_{\rm eff}
    = \frac{i}{(2\pi)^3}\,\ln\frac{1}{b}
    \int\limits_{\bm{x}\bm{y}\bm{z}}&
    \mathcal{K}_{\bm{x}\bm{y}\bm{z}}\,
    \big[\rho_{\infty}^a(\bm{x})
    \rho_{\infty}^a(\bm{y})
    +\rho_{-\infty}^a(\bm{x})
    \rho_{-\infty}^a(\bm{y})
    \nonumber\\
    &-\rho_{\infty}^a(\bm{x})
    W_{ab}(\bm{z})
    \rho_{-\infty}^b(\bm{y})
    -\rho_{-\infty}^a(\bm{x})
    W^{\dagger}_{ab}(\bm{z})
    \rho_{\infty}^b(\bm{y}) \big],
\end{align}
with the transverse kernel $\mathcal{K}_{\bm{x}\bm{y}\bm{z}}$ as
defined in Eq.~(\ref{Kdef}). The associated evolution Hamiltonian
$H_{\rm BREM}$ is then obtained by replacing the color fields
$\dA^-_a$ in the temporal Wilson lines by functional derivatives
with respect to the color source $\rho_a(x^+,\x)\equiv\int \dif
x^- \rho_a(x)$, cf. Eq.~(\ref{atorhoBREM}), which yields
 \be\label{Wbrem}
W(\bm{x})\,=\,
 {{\rm T}} \exp\left(- g \int
    \dif x^+ \frac{\delta}{\delta \rho (x^+,\bm{x})}\right),
    \ee
The ensuing Hamiltonian has the structure anticipated in
Eq.~(\ref{BREM0}) and describes brems--\\strahlung in the BFKL
evolution of a dilute target (cf. Fig. \ref{JBfig}.b). As
originally observed in Ref. \cite{KL3}, $H_{\rm BREM}$ is dual to
the JIMWLK Hamiltonian (\ref{HJIMWLK}) in the sense of the
transformation (\ref{DUAL}). As explained in Sect. \ref{S_SEFF},
this duality reflects the fact that the same diagram, namely Fig.
\ref{JBfig}.b, can describe bremsstrahlung with decreasing $k^+$
in the target wavefunction, or gluon merging with decreasing $k^-$
in the wavefunction projectile.

In fact, this duality refers to the complete Hamiltonian
structure, including the Poisson brackets. Specifically, by using
Eq.~(\ref{Wbrem}), one can deduce the commutation relations
 \be\label{commutrW}
 \big[\rho^a_\infty(\bm{x}), \,
W_{bc}(\bm{y})\big]&=&g(T^aW(\bm{x}))_{bc}\,\delta^{(2)}({\bm{x-y}})
\nn \big[\rho^a_{-\infty} (\bm{x}), \,
W_{bc}(\bm{y})\big]&=&g(W(\bm{x})T^a)_{bc}\,\delta^{(2)}({\bm{x-y}}),
 \ee
which are similar to Eq.~(\ref{LieFin}) and show that the color
charges $\rho^a_{\pm\infty}$ act as infinitesimal gauge rotations
of the Wilson line $W$ at its end points (or, equivalently, as
functional derivatives w.r.t. the field $\dA^-_a$ at
$x^+=\pm\infty$). This in turn implies that, within the
Hamiltonian theory defined by $H_{\rm BREM}$, $\rho^a_\infty$ and
$\rho^a_{-\infty}$ must be treated as {\em non--commuting
variables}, which provide two independent representations of the
SU$(N_c)$ color algebra :
 \be\label{commrho}
 \big [\rho^a_\infty(\x),\, \rho_\infty^b(\y) \big]
  &\,=\,& - if^{abc}\rho_\infty^c(\x) \,\delta^{(2)}(\x-\y)\,,\nn
 \big [\rho^a_{-\infty}(\x),\,
 \rho_{-\infty}^b(\y) \big] &\,=\,&  if^{abc}\rho_{-\infty}^c(\x)
 \,\delta^{(2)}(\x-\y)\,, \nn
 \big[\rho^a_\infty(\x), \, \rho_{-\infty}^b(\y)\big] &\,=\,& 0 .\ee
Like in the JIMWLK case, the commutators (\ref{commrho}) are
enforced by Eq.~(\ref{commutrW}) together with the Jacobi identity
and  the following relation (cf. Eq.~(\ref{rhoinfdef}))
 \be\label{rhoend}
 \rho^a_\infty(\x) \,=\,W^{ab}(\bm{x})\,
 \rho_{-\infty}^b({\x}),
 \ee
showing that the color charge densities at the `end points'
$x^+=\infty$ and $x^+=-\infty$ are not independent quantities, as
they are related by a gauge rotation.

We recognize in the equations above the `dual' version of the
commutators (\ref{LieFin})--(\ref{commLR}) for the JIMWLK problem,
with the precise duality transformation being now:
 \be\label{DUALLIM}\hspace*{-0.5cm}
 J_L^a(\x)\,\longleftrightarrow \,\rho^a_\infty(\x)
 \,,\qquad
 J_R^a(\x)\,\longleftrightarrow \,\rho_{-\infty}^a(\x)
 \,,\qquad
 V^\dagger(\x)\longleftrightarrow \,W(\x)
 \,.\ee
In particular, the relation (\ref{rhoend}) between $\rho^a_\infty$
and $\rho^a_{-\infty}$ is dual to the relation (\ref{JLRdef})
between $J^a_L$ and $J^a_R$, and they both emerge as particular
limits of the more general relation (\ref{diamond}). Still as in
the JIMWLK problem, it becomes advantageous to use
Eq.~(\ref{rhoend}) to eliminate $\rho^a_{-\infty}$ from the
problem and thus arrive at the following Hamiltonian
 \begin{align}\label{Slow2}
    H_{\rm BREM}
    = \frac{1}{(2\pi)^3}\,
    \int\limits_{\bm{x}\bm{y}\bm{z}}
    \mathcal{K}_{\bm{x}\bm{y}\bm{z}}\,
    \rho^a_{\infty}(\bm{x})
    \left[1 +
    W_{\bm{x}}
    W^{\dagger}_{\bm{y}}-
    W_{\bm{x}}
    W^{\dagger}_{\bm{z}}-
    W_{\bm{z}}
    W^{\dagger}_{\bm{y}} \right]^{ab}
    \rho^b_{\infty}(\bm{y}),
\end{align}
which involves only two independent variables, $\rho^a_\infty$ and
$W_{ab}=W^\dagger_{ba}$, and is manifestly dual to the JIMWLK
Hamiltonian (\ref{HJIM}). The evolution equation for an arbitrary
observable $\mathcal{O}(\rho_\infty,W)$ is then obtained as
    \be\label{HO}
    \frac{\del}{\del \tau}\lan \mathcal{O} \ran_{\tau}
    = -\big\lan \big[H_{\rm BREM},\mathcal{O}\,
     \big] \big\ran_{\tau},
    \ee
which is the analog of Eq.~(\ref{EVOLCAN}) in the JIMWLK case.

By virtue of the duality property it is furthermore clear that the
BREM Hamiltonian endowed with the above commutation relations acts
on the temporal  Wilson lines in the same way as the JIMWLK
Hamiltonian (\ref{HJIM}) acts on the longitudinal Wilson lines.
Accordingly, the evolution equations generated by $H_{\rm BREM}$
for the operators built with $W$ and $W^\dagger$ (e.g., ${\rm Tr}
W(\bm{x}) W^\dagger(\bm{y})$) are formally identical to the
Balitsky equations \cite{B} obeyed by the corresponding operators
built with $V$ and $V^\dagger$. This should not come as a
surprise: when focusing on the correlators of $W$, we are viewing
the diagram in  Fig. \ref{JBfig}.b upside down, as the
JIMWLK--like evolution of the projectile (our left mover). Then a
2--point function like ${\rm Tr} W(\bm{x}) W^\dagger(\bm{y})$
describes the scattering between a right--moving dipole and the
left--moving system which is evolving.

But, of course, our main interest when using Eq.~(\ref{Slow}) is
not in describing the JIMWLK evolution of a dense projectile, but
rather the fluctuations associated with bremsstrahlung in the
evolution of a dilute target. These fluctuations can be probed via
scattering with a projectile which is more complex than a simple
dipole (since the projectile must measure a higher--point
correlation function of the gluon fields in the target). Consider,
for instance, a projectile made with two dipoles. According to
Sect. \ref{S_SCATT}, the amplitude for the simultaneous scattering
of the two dipoles in the dilute regime is computed as:
 \be\label{T2dipole_weak} \langle T^{(2)}(\x_1,\y_1;\x_2,\y_2)
\rangle_\tau \, \simeq \,
  \frac{g^4}{16N_c^2}\,\big \langle
 \big(\alpha_a(\x_1)-\alpha_a(\y_1)\big)^2\,
 \big (\alpha_b(\x_2)-\alpha_b(\y_2)\big)^2
 \big\rangle_\tau\,,\ee
where $\alpha^a(\x)\equiv \int\dif x^- \alpha^a_\infty(x^-,\x)$ is
the color field in the target at the time of scattering, and is
related to $\rho^a_{\infty}(\bm{x})$ via Eq.~(\ref{alpha}) :
$\alpha^a(\x)= \int d^2\y\,
\Delta(\x-\y)\,\rho^a_{\infty}(\bm{y})$.  Note however a subtle
point concerning the definition (\ref{T2dipole_weak}) of $\langle
T^{(2)} \rangle_\tau$ : Since the fields $\alpha^a(\x)$ do not
commute with themselves, the relative ordering of the individual
dipole operators within Eq.~(\ref{T2dipole_weak}) turns out to be
important : $\langle  T(1) T(2)\rangle_\tau\ne \langle  T(2)
T(1)\rangle_\tau$ ! It remains as an interesting problem to
clarify the physical meaning of this potential ambiguity, and
construct the evolution equation for $\langle T^{(2)}
\rangle_\tau$ according to Eq.~(\ref{HO}).

But such ordering ambiguities are expected to disappear, at least,
in the large--$N_c$ limit, where the color dipole picture is valid
\cite{AM94,AM95,IM031} and the evolution equations for the dipolar
scattering amplitudes have been recently constructed
\cite{IT04,MSW05,IT05,BIIT05}. In this limit,  and for
sufficiently large energy, the evolution generated by $H_{\rm
BREM}$ should reduce to the equations in Refs.
\cite{IT05,MSW05,BIIT05}.

Furthermore, all such ambiguities  trivially disappear in the BFKL
approximation in which the fluctuations are completely neglected
(meaning that the Hamiltonian is evaluated to second order in the
functional derivatives). To deduce the BFKL Hamiltonian from the
bremsstrahlung one, it is convenient to first rewrite the
expression within  the square brackets in Eq.~(\ref{Slow2}) as
$(1-W_{\bm{x}} W^{\dagger}_{\bm{z}})(1-W_{\bm{z}}
W^{\dagger}_{\bm{y}})$. Then to the order of interest it is enough
to expand the various Wilson lines to first order:
\begin{align}\label{Wilsone}
    W^{ab}(\bm{x})\approx \delta^{ab} + g f^{abc}
    \int_{-\infty}^{\infty} \dif x^+ \dA^-_c (x^+,\bm{x}).
\end{align}
The BFKL Hamiltonian is finally obtained by using the identification
(\ref{atorhoBREM}) and neglecting the time--dependence of $\rho$ :
\begin{align}\label{SBFKL}
    H_{\rm BFKL} =
    \frac{ g^2}{(2\pi)^3}
    \int\limits_{\bm{x}\bm{y}\bm{z}}&
    \mathcal{K}_{\bm{x}\bm{y}\bm{z}}\,
    f^{ace}f^{bde}
    \rho^a(\bm{x})
    \left[ \frac{\delta}{\delta \rho^c(\bm{x})} -
           \frac{\delta}{\delta \rho^c(\bm{z})} \right]
    \left[ \frac{\delta}{\delta \rho^d(\bm{z})} -
           \frac{\delta}{\delta \rho^d(\bm{y})} \right]
        \rho^b(\bm{y}).
\end{align}
This is recognized as the expression of the BFKL Hamiltonian which
emerges naturally in applications of the CGC formalism to the
dipole picture \cite{IM031,MSW05,BIIT05}. It is furthermore dual
to the corresponding expression obtained from the appropriate
expansion of the JIMWLK Hamiltonian (\ref{HJIM})
\cite{IM031,ODDERON}.

\section{Conclusions}

In this paper we have constructed the effective Hamiltonian
describing the evolution of gluon correlations in a hadron
wavefunction with increasing energy in the leading logarithmic
approximation. In its essence, this Hamiltonian describes BFKL
evolution in the presence of gluon recombination and
bremsstrahlung, and thus of `Pomeron loops'. It thus provides the
appropriate framework to follow the evolution of a hadronic system
all the way up in energy, from a dilute initial state up to a high
density state characterized by gluon saturation and the formation
of a color glass condensate.

A priori, the effective Hamiltonian governs the evolution of a
single hadron wavefunction (say, the target in a high--energy
collision), but it can be also applied to scattering once a
factorization scheme is known (as is the case, for instance, for a
simple projectile built with dipoles). The Hamiltonian has the
fundamental property of self--duality, which reflects boost
invariance and suggests that the present formalism should
naturally allow for a symmetric description of the evolution in
the high--energy scattering.

The Hamiltonian has been obtained within a renormalization group
analysis which exploits the separation of scales in rapidity and
thus entails a coarse--graining in both the longitudinal and the
temporal directions. The RG analysis is successful because the
separation of scales is preserved by the evolution: the
interactions responsible for the evolution are localized within a
space--time region which is small on the resolution scale of the
quantum gluons which are integrated out in a single step of the
evolution. Because of that, and of multiple scattering, the
effective action describing one--step evolution involves the
target fields only through Wilson lines. More precisely, it
involves three independent group--valued, two--dimensional, field
degrees of freedom, which are any three among the four Wilson
lines forming a Wilson loop with unit trace around the interaction
region. Two of these Wilson lines are path--ordered in the
longitudinal direction and describe the recombination processes
responsible for saturation and unitarity corrections. The other
two are ordered in time and describe the bremsstrahlung processes
responsible for gluon number fluctuations and the associated
correlations. Both type of processes play an essential role in the
evolution towards high gluon density with increasing energy.

The present construction encompasses and generalizes previous
approaches in the literature, like the effective action approaches
in Refs. \cite{VV93,LKSeff,LipatovS,B1}, the Balitsky--JIMWLK
equations \cite{B,JKLW97,RGE,W}, the recently derived evolution
equations with Pomeron loops (as valid at large $N_c$)
\cite{IT04,IT05,MSW05,LL05,BIIT05}, and the effective Hamiltonian
for the dilute regime by Kovner and Lublinsky \cite{KL05,KL3}. In
most cases, the correspondence with such previous results has been
explicitly demonstrated in the appropriate limits. For instance,
in the situation where one of the background fields is weak, our
effective action reduces to that derived by Balitsky \cite{B1} for
asymmetric scattering. It turn, this situation encompasses two
different physical regimes: the high--density regime (the radiated
fields are weak) where the JIMWLK Hamiltonian \cite{RGE,W} has
been recovered, and the dilute regime (the target fields are
weak), where the BREM Hamiltonian \cite{KL3} has been shown to
emerge.

But some other correspondences are still to be explored. Our
general action, as given by any of the equations (\ref{SVW}),
(\ref{selfdual}), (\ref{selfdual1}) or (\ref{SVIW}), does not look
similar to the action obtained in Ref. \cite{B1} for the collision
between two high--density hadrons. Also, the relation to the
effective action for reggeized gluons by Lipatov and collaborators
\cite{LKSeff,LipatovS} remains unclear to us. Finally, it would be
interesting to extract the large--$N_c$ limit of our present
results and thus verify whether this is equivalent to the Pomeron
effective theory of Refs. \cite{IT05,BIIT05}, as {\em a priori}
expected.

Another important open problem refers to the understanding of the
Hamiltonian structure of the effective theory in the general case.
The Poisson brackets that would be most naively written down ---
by the straightforward generalization of the corresponding
analysis in the two (JIMWLK and BREM) limiting cases --- would
spoil the two--dimensional nature of the effective action (in the
sense of restoring the longitudinal and temporal coordinates), and
thus are unacceptable. Other `quantization' prescriptions, like
the construction of a path integral in rapidity (so like in Refs.
\cite{PATH,BalitskyPath}), are currently under investigation.

\vfill 
\section*{Acknowledgments}

The authors would like to thank Jean--Paul Blaizot, Kazu Itakura,
and Al Mueller for many insightful conversations.  Y. Hatta and L.
McLerran acknowledge inspiring and imaginative discussions with
Alex Kovner and Misha Lublinsky. Y. H. is supported by Special
Postdoctoral Research Program of RIKEN. This manuscript has been
authorized under Contract No. DE-AC02-98CH10886 with the U. S.
Department of Energy. This research has been partially supported
by the Polish Committee for Scientific Research, KBN Grant No. 1
P03B 028 28.


\end{document}